\documentclass{aa}  

\usepackage{graphicx}
\usepackage{txfonts}
\usepackage{lipsum}
\usepackage{subcaption}
\usepackage{lscape}             
\usepackage{placeins}           

\usepackage[colorlinks=true, linkcolor=blue, citecolor=blue, urlcolor=blue]{hyperref}

\begin{document}


\title{Tracing expansion-driven star formation in the Perseus molecular cloud using young stellar objects}


%
%
%

\author{Chen-Rui Gu\inst{1}
\and Zhi-Kai Zhu\inst{2,3}\email{}
\and Zu-Jia Lu\inst{1,4}\email{luzujia@gxu.edu.cn}\corrauth{luzujia@gxu.edu.cn}        
\and Min Fang\inst{2}\email{mfang@pmo.ac.cn}
\and Longhui Yang\inst{1}\email{}
\and Cheng-Yang He\inst{1}\email{hecyhaibara@163.com}}

\institute{Guangxi Key Laboratory for Relativistic Astrophysics, Department of Physics, Guangxi University, Nanning 530004, China
\and Purple Mountain Observatory, Chinese Academy of Sciences, 10 Yuanhua Road, Nanjing 210023, China
\and School of Astronomy and Space Science, University of Science and Technology of China, Hefei 230026, China
\and Department of Physics, University of Oxford, Keble Road, Oxford OX1 3RH, UK}

\date{Received July XX, 2026}

 
\abstract
{Recent studies have increasingly linked star formation in the solar neighbourhood to the expansion of the Local Bubble. The Per-Tau Shell (PTS), a vast 3D dust structure hosting the Perseus and Taurus clouds on its surface, is a candidate for such triggered formation. While the Local Bubble primarily drives star formation in Taurus, Perseus is kinematically distinct from this flow. Consequently, the dynamical influence of the PTS on the Perseus cloud remains an open question.}
{We aim to reconstruct the star formation history of the Perseus molecular cloud and investigate its potential link to the historical expansion of the PTS.} 
{We performed a 6D phase-space analysis of 406 young stellar objects (YSOs) of Perseus. Our approach integrates Gaia DR3 kinematics with Bayesian age inference and a heteroscedastic likelihood framework to resolve radial gradients in YSO ages and intrinsic tangential velocity dispersions relative to the PTS centre. Additionally, we constructed a unified catalogue by integrating existing infrared-based classifications from the literature to examine the projected spatial segregation of different evolutionary stages and compared stellar 3D kinematics (proper motions and line-of-sight velocities) with $^{12}$CO and $^{13}$CO gas velocity fields to assess kinematic coherence.}
{The YSOs exhibit coherent outward motions relative to the centre of the PTS, a negative age-distance gradient with younger populations at larger radii, and a radial decline in tangential velocity dispersions. Younger YSO classes also show a stronger directional concentration towards the PTS centre. Furthermore, both the coherent proper motions and the line-of-sight velocity gradients of these YSOs align with the gas velocity field, indicating large-scale kinematic coherence.}
{The Perseus YSOs exhibit concordant kinematic and temporal signatures consistent with an expansion-driven origin. These results provide robust evidence for a layered, inside-out pattern of star formation, suggesting that the historical expansion of the PTS triggered a sequential progression of star birth and effectively shaped the star formation history of the Perseus complex.}

\keywords{{Star formation}{1569} -- {Young stellar objects}{1834} -- {Stellar kinematics}{1608} -- {Interstellar clouds}{834}}

\authorrunning{Gu}
\titlerunning{Expansion-driven star formation in Perseus}
\maketitle\nolinenumbers


\section{Introduction}

Star formation occurs within molecular clouds that are shaped by the interplay of turbulence \citep{Hennebelle+2012A&ARv, Guerrero-Gamboa+2020ApJ,Lu+2020ApJ}, magnetic fields \citep{Ostriker+2001ApJ, McKee+2007ARA&A}, self-gravity, and stellar feedback \citep{Lada+2003ARA&A, Rey-Raposo+2017MNRAS}. Recent observations reveal that many prominent star-forming regions in the solar neighbourhood are embedded within or organized along large-scale interstellar structures \citep{Lallement+2014, Alves+2020}. Among the key drivers shaping these environments are expanding superbubbles sculpted by stellar and supernova feedback. A well-documented example is the Local Bubble, which has been demonstrated to trigger star formation via an expansion-driven mechanism, sweeping up the interstellar medium to initiate subsequent cloud formation and star birth \citep{Zucker+2022Nature}. In parallel, recent 3D dust mapping has revealed the existence of the Per-Tau Shell (PTS), another vast, nearly spherical structure that hosts the Perseus and Taurus molecular clouds on its surface \citep{Bialy+2021ApJL, Edenhofer+2024AandA, ONeill+2024ApJ}. However, while the Local Bubble has been extensively investigated, research on the PTS remains scarce, with current knowledge largely confined to static dust morphology. Consequently, our understanding of its potential role as an expansion-driven trigger and the associated kinematic imprints on the enclosed stellar populations is still highly limited, necessitating detailed dynamical investigations.

Young stellar objects (YSOs) are ideal dynamical tracers for star formation, as they inherit and preserve the initial kinematic properties of their natal molecular clouds. Extensive infrared surveys have historically provided the basis for the large-scale identification and classification of these sources. Based on this, recent studies using Gaia's astrometric precision have identified a large population of YSOs in the Perseus complex \citep{Ortiz-Leon+2018ApJ, Pavlidou+2021MNRAS, Kounkel+2022AJ, Wang+2022ApJ}. Gaia data release 3 (DR3) provides high-precision astrometry and 3D motions for large samples of YSOs \citep{Gaia+2023}. These new data have enabled YSOs to serve as dynamical feedback tracers in diverse environments \citep{Yang+2025ApJS}. Furthermore, they allow us to track the assembly and dispersal of both clustered and distributed populations in the solar neighbourhood and to relate these YSO populations to the surrounding large-scale interstellar structures \citep{Zhu+2024ApJ}. Leveraging these advancements, the Perseus molecular cloud, residing on the PTS surface and exhibiting complex kinematics distinct from the Local Bubble flow, presents an ideal laboratory for investigating the relationship between young stellar populations and large-scale structures. 

The Perseus star-forming region has been extensively observed, yielding a coherent view of large-scale cloud morphology and magnetic fields as well as detailed kinematic studies of small-scale filaments, cores, and YSOs. Radial-velocity observations show that dense cores in Perseus have only small velocity offsets from their surrounding envelopes, suggesting that they are largely co-moving with the parent cloud \citep{Kirk+2010ApJ}. Proper-motion studies further indicate that YSOs in the main subgroups exhibit mostly random residual motions, without clear evidence for internal expansion \citep{Ortiz-Leon+2018ApJ}. Regarding the region's formation history, an atypical cloud–cloud collision has been proposed as a trigger for star formation in Perseus \citep{Kounkel+2022AJ}. Within the local Galactic architecture, Perseus resides off the surface of the Local Bubble and is a dynamical outlier in the Radcliffe wave, pointing to a distinct physical driver for its observed kinematics \citep{Konietzka+2024Nature}. Despite extensive observational efforts in Perseus, it remains unclear whether the historical expansion of the PTS triggered star formation in the Perseus molecular cloud. While recent 3D dust models place the Perseus molecular cloud on the surface of the PTS, research on the PTS itself is still in its infancy compared to the well-studied Local Bubble.

To date, the link between the PTS and Perseus has relied primarily on static spatial overlap, lacking dynamical confirmation. In this study, we investigated this physical link by quantifying the radial gradients in stellar ages and intrinsic velocity dispersions. Our approach goes beyond simple morphology, utilizing YSOs as dynamic tracers to reconstruct the expansion history of the shell. We analysed the 3D positions and velocities of 406 YSOs in a uniform heliocentric-Galactic frame. The radial trends in stellar ages and intrinsic tangential velocity dispersions were established through a sliding-window analysis centred on the PTS. As an independent projected cross-check, we used infrared YSO classifications to examine class-dependent angular separations from the PTS-centre direction. Furthermore, we evaluated the dynamical link between the stellar populations and their natal environment by comparing the 3D stellar kinematics with the $^{12}$CO and $^{13}$CO gas velocity fields. Through these analyses, we aimed to gain a more comprehensive understanding of the relationship between Perseus YSOs and the PTS. Using Perseus YSOs as tracers, this study provides dynamical constraints to evaluate the expansion-driven star formation scenario.

The structure of the paper is as follows. Section \ref{data_methods} details the datasets utilized, the criteria applied for sample selection, and the analytical methodology used throughout the work. The results derived from our analysis are presented in Sect. \ref{results}. The implications of our results are discussed in Sect. \ref{discussion}, and the main conclusions are summarized in Sect. \ref{conclusions}.

\section{Data and methods}
\label{data_methods}

\subsection{YSOs selection and preprocessing}
\label{YSOs selection and preprocessing}
In this work, we adopted our primary 3D kinematic sample from the catalogue presented in \citet{Olivares+2023AandA}, the most recent and comprehensive census of YSOs in Perseus, which also provides estimates of extinction per-source, $A_V$. Details of that catalogue can be found in the paper and are brieﬂy summarized here. In their catalogue, the full Cartesian velocity components in the International Celestial Reference System (ICRS) ($U, V, W$) intrinsically rely on the radial velocities compiled by cross-matching members with the APOGEE survey \citep{Abdurrouf+2022ApJS}, Gaia DR3 \citep{Gaia+2023}, and the SIMBAD database \citep{Wenger+2000A&AS}. They adopted APOGEE radial velocities with high precision, if available. Otherwise, they used the values of Gaia DR3 or SIMBAD. The \citet{Olivares+2023AandA} catalogue covers a sky region of $51^\circ < \mathrm{RA} < 59^\circ$ and $30^\circ < \mathrm{Dec} < 33^\circ$ and consequently does not include the extended Perseus subgroups Electryon, Mestor, and Cynurus. The physical implications of this exclusion are discussed in Sect.~\ref{discussion_sample}.

To ensure uniform photometric and astrometric quality, we cross-matched each source with Gaia DR3 by source ID and attached key quality indicators. These include broadband photometry \(\mathit{G},\, \mathit{G}_{\mathrm{BP}},\, \mathit{G}_{\mathrm{RP}}\) with uncertainties \(\sigma_{\mathrm{G}},\, \sigma_{\mathrm{BP}},\, \sigma_{\mathrm{RP}}\), and the re-normalized unit weight error \textit{RUWE}. We denote the parallax-based signal-to-noise ratio by \(\mathrm{S/N}\) and define it as \(\mathrm{S/N} \equiv \varpi/\sigma_{\varpi}\), where \(\varpi\) is the parallax and \(\sigma_{\varpi}\) is its uncertainty. We then applied three criteria to select a clean and well-measured sample. (i) We required a $\mathrm{S/N} = \varpi/\sigma_{\varpi} \ge 5$, which retains stars with reliable parallaxes and distances. (ii) We required $\sigma_{\mathrm{G}} \le 0.054\,\mathrm{mag}$, $\sigma_{\mathrm{BP}} \le 0.054\,\mathrm{mag}$, and $\sigma_{\mathrm{RP}} \le 0.054\,\mathrm{mag}$, which ensures high-quality G, BP, and RP photometry with uncertainties at the level of about five percent in flux. (iii) We required an $\textit{RUWE} \le 1.4$ and non-missing, which removes sources with poor or non-single-star astrometric solutions \citep{Lindegren+2021}. To remove the systematic motions introduced by the solar peculiar velocity, we transformed the heliocentric 3D velocities into the Local Standard of Rest (LSR) frame. We adopted the peculiar solar motion of $(U, V, W)_{\odot} = (11.10, 12.24, 7.25)$~km ~ s$^{-1}$ \citep{Schonrich+2010MNRAS}. All subsequent kinematic analyses and vector maps are presented in this LSR frame.

To establish the evolutionary stages of our sample, we collated YSO classifications from infrared datasets published in several key studies presented in \citet{Muench+2007AJ, Gutermuth+2008ApJ, Enoch+2009ApJ, Evans+2009ApJS, Gutermuth+2009ApJS, Kryukova+2012AJ, Dunham+2015ApJS, Young+2015AJ, Tobin+2016ApJ, Marton+2016MNRAS, Mercimek+2017AJ, RuizRodriguez+2018MNRAS, Lalchand+2022AJ}. These individual classifications were integrated into a unified catalogue using a 2 arcsec spatial cross-matching radius to remove duplicate entries. This is an angular criterion applied among the infrared classification catalogues on the sky plane to unify evolutionary-stage labels. It is independent of any assumed distance and does not enter the kinematic sample selection. At the characteristic distance of the Perseus complex ($\sim$300--350 pc), this 2 arcsec threshold corresponds to a projected physical separation of $\sim$600--700 au, which is well below the typical spacing of distinct YSOs in the region. This criterion is sufficiently small to confidently merge multiple catalogue entries corresponding to the same physical source while mitigating the spurious merging of genuinely distinct, nearby YSOs.

Based on the selection criteria above, we established a kinematic sample of 406 YSOs with 6D phase-space data summarized in Table~\ref{tab:mr-perseus-pts} of Appendix~\ref{app_406YSOs}. These 406 sources are distributed among six kinematic groups identified by \citet{Olivares+2023AandA}: IC\,348 (160), Gorgophone (77), Alcaeus (68), Heleus (61), NGC\,1333 (27), and Autochthe (13). Separately, the infrared data compiled via the cross-matching approach are summarized in Table \ref{tab:ir-yso-sample} of Appendix~\ref{app_infrared_YSO}.

\begin{figure}
\centering
\includegraphics[width=\hsize]{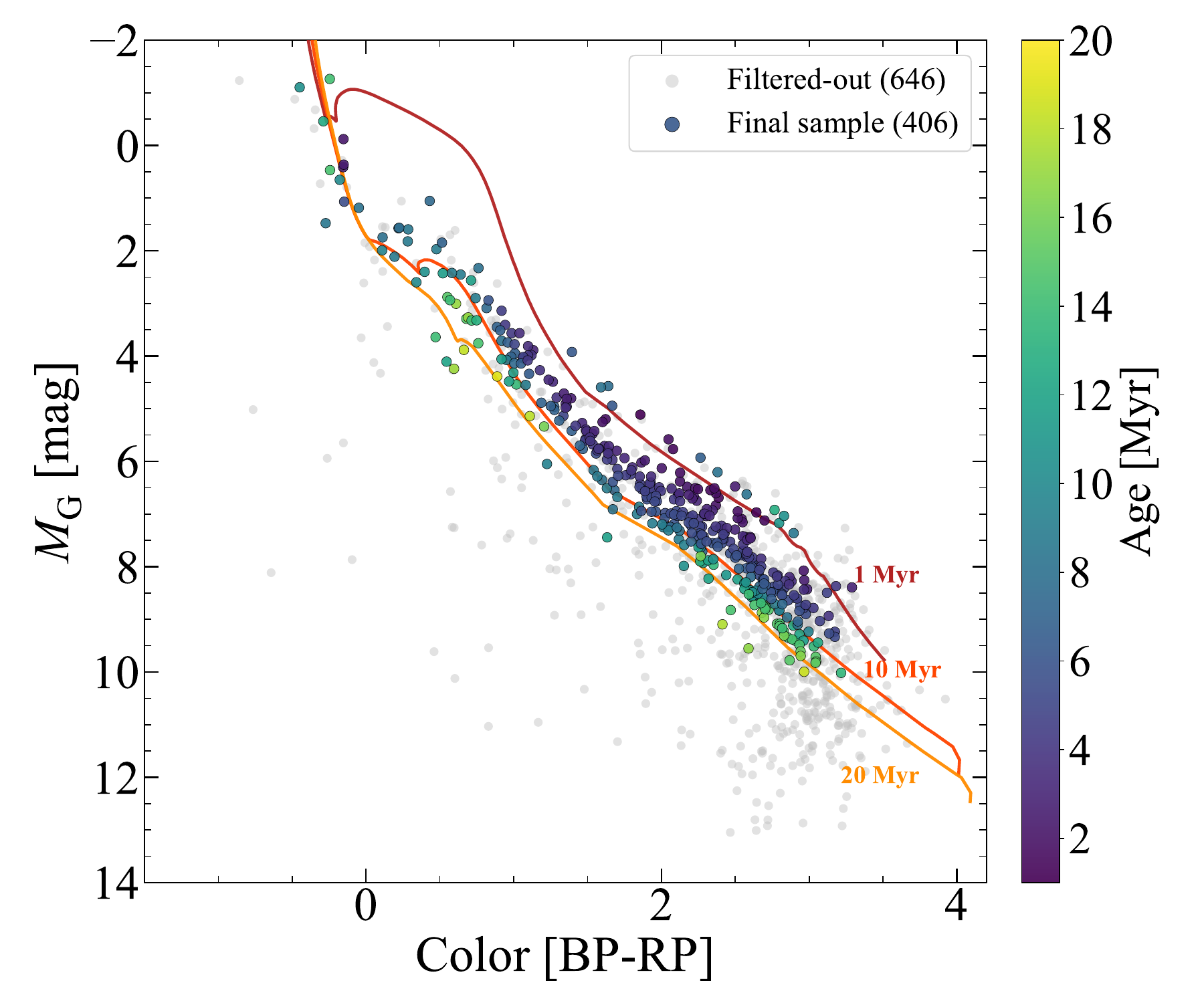}
\caption{CMD of the Perseus sample. The grey points show sources from the parent catalogue of \citet{Olivares+2023AandA} that do not satisfy the quality criteria (i)--(iii) described in Sect.~\ref{YSOs selection and preprocessing} (parallax $\mathrm{S/N} \ge 5$, photometric uncertainties $\le 0.054$~mag, and $\mathit{RUWE} \le 1.4$). The coloured points represent the final kinematic sample of 406 YSOs, colour-coded by their posterior median age. Three representative PARSEC isochrones at 1, 10, and 20~Myr are overplotted as solid lines for reference.}
\label{cmd}
\end{figure}

\subsection{Bayesian inference of YSO ages}

Prior to age inference, we applied a first-order de-reddening to the Gaia bands using $A_V$.
Specifically, we corrected for the magnitudes $G$/BP/RP with fixed coefficients
$(k_{\mathrm{G}},k_{\mathrm{BP}},k_{\mathrm{RP}})=(0.789,\,1.002,\,0.589)$, consistent with an average
Milky Way extinction law ($R_{\mathrm{V}}\simeq3.16$) \citep{Wang+2019ApJ}.
The extinction-corrected magnitudes of BP and RP define the colour, and the extinction-corrected magnitude $G$, together with the per-source heliocentric distance $d \equiv 1/\varpi$, defines $M_{\mathrm{G}}$. Explicitly,
\begin{equation}
M_{\mathrm{G}} = G - k_{\mathrm{G}}\,A_V - 5\,\log\!\left(\frac{d}{10\;\mathrm{pc}}\right) \, .
\end{equation}
The associated uncertainty is
\begin{equation}
\sigma_{M_{\mathrm{G}}}^2 = \sigma_G^2 + \left(\frac{5}{\ln 10}\,\frac{\sigma_d}{d}\right)^2 + \left(k_{\mathrm{G}}\,\sigma_{A_V}\right)^2 \, ,
\end{equation}
where $\sigma_\mathrm{d}$ is derived from the parallax uncertainty via $\sigma_\mathrm{d} = d^2\,\sigma_\varpi$, and $\sigma_{A_V}$ is estimated from the 95\% confidence interval of the extinction posterior in \citet{Olivares+2023AandA}. We based our age analysis on the absolute magnitude of the $G$ band. Although the $G_{\rm RP}$ band is often used for YSOs, we verified that in our sample the $G$-band photometric errors are systematically smaller than those in $G_{\rm RP}$, making $M_{\mathrm{G}}$ the most precise photometric input for Bayesian inference.

We estimated the YSO ages by fitting their positions on the colour-magnitude diagram (CMD) to PARSEC theoretical isochrones, as shown in Fig.~\ref{cmd} \citep{Bressan+2012MNRAS}. Ages were inferred per star by evaluating a 1D Gaussian likelihood in absolute magnitude $M_{\mathrm{G}}$ at a given colour $(\mathrm{BP{-}RP})$ on a fixed age grid $a \in [1,20]\,\mathrm{Myr}$ with a $0.25\,\mathrm{Myr}$ step. Here, we denote the Gaia colour by $C\equiv (\mathrm{BP{-}RP})$. The quantity $M_{\rm G}^{\rm iso}(a,C)$ is the PARSEC isochrone-predicted absolute magnitude at age $a$ and colour $C$, obtained by interpolation on the model grid. The uncertainty $\sigma_\mathrm{C}$ is the photometric error in $C$, and the local slope $\frac{dM_{\rm G}^{\rm iso}}{dC}$ was evaluated along the isochrone. We define $\sigma_{M_{\rm G}}$ as the uncertainty of $M_{\rm G}$, obtained by propagating the uncertainties in the apparent $G$-band magnitude, the heliocentric distance, and the extinction correction. Assuming that these contributions are independent, we added them in quadrature. Moreover, a model dispersion term of $\sigma_{\mathrm{m}} = 0.1$ mag was incorporated into the likelihood calculation to account for systematic model uncertainties \citep{Liu+2025SCPMA}. The resulting total uncertainty is given by
\begin{equation}
\sigma_{\rm tot}^2
= \sigma_{M_{\rm G}}^2 + \sigma_{\rm m}^2
+ \left(\frac{dM_{\rm G}^{\rm iso}}{dC}\,\sigma_\mathrm{C}\right)^2 \,.
\end{equation}
The corresponding log-likelihood for each star is
\begin{multline}
\ln \mathcal{L}(M_{\mathrm{G}} \mid a,\mathrm{C})
= -\frac{1}{2}\Biggl[
  \frac{(M_{\mathrm{G}} - M_{\mathrm{G}}^{\mathrm{iso}}(a,\mathrm{C}))^2}{\sigma_{\rm tot}^2}
\\
 + \ln(2\pi) + \ln\!\big(\sigma_{\rm tot}^2)
\Biggr] \,.
\end{multline}

\begin{figure}
\centering
\includegraphics[width=\hsize]{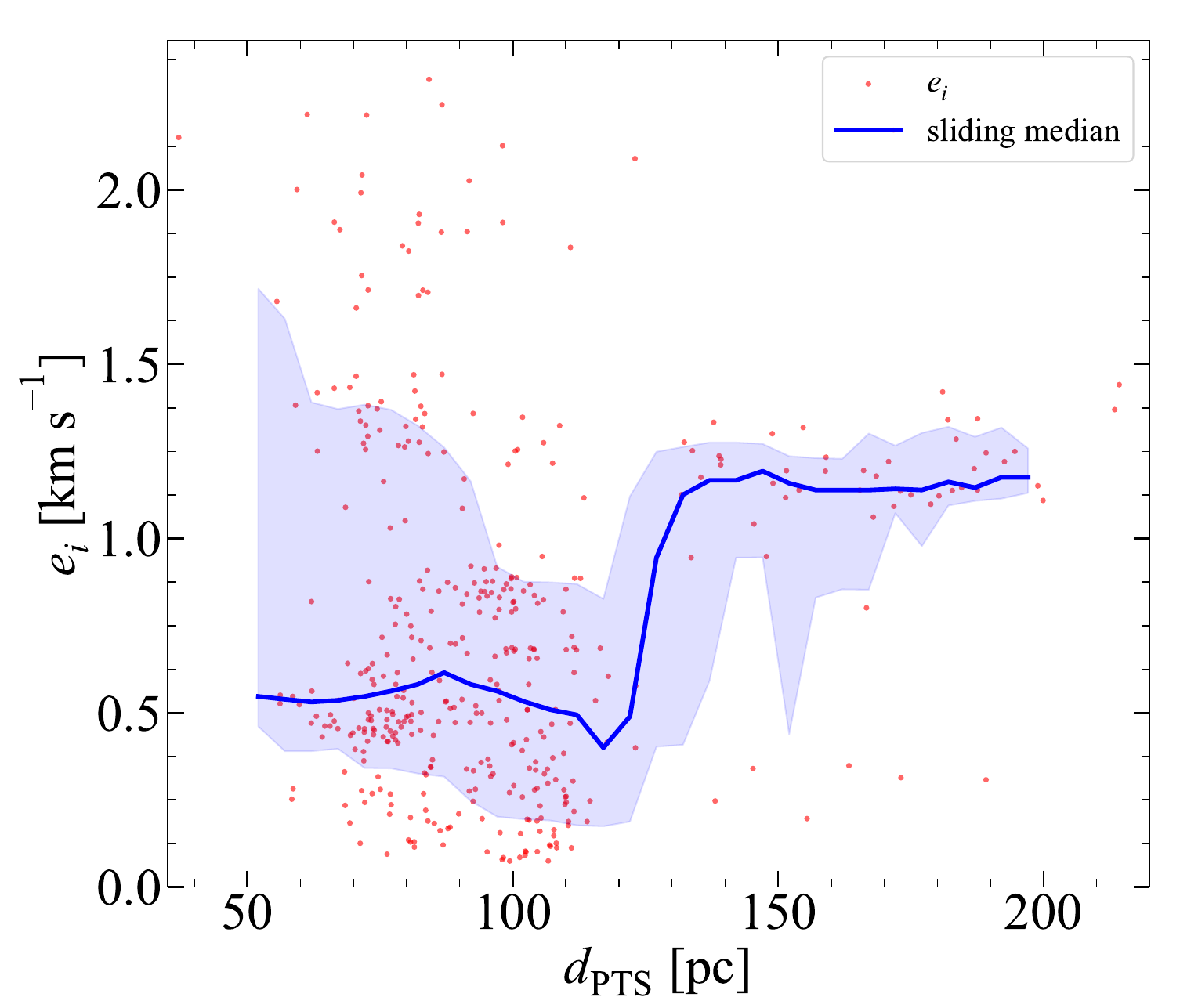}
\caption{Measurement error of tangential-velocity of each star $e_i = \sqrt{e_{{\rm RA}^\ast}^2 + e_{\mathrm{Dec}}^2}$ as a function of 3D distance to the PTS centre. The red points show individual stars, and the blue curve and shaded region denote the sliding-window median and 16th--84th percentile envelope, respectively. The increase in $e_i$ with distance motivates the use of heteroscedastic MLE adopted in this work.}
\label{fig:error_vs_distance}
\end{figure}

Using extinction-corrected colours and absolute magnitudes, we computed an age posterior for each star on a fixed grid with a uniform prior in age and summarized it by the posterior median $t_{\mathrm{med}}$. To calculate the 3D distance of each YSO from the PTS centre $d_{\mathrm{PTS}}$, we adopted the Heliocentric Galactic Cartesian coordinates $(X, Y, Z) = (-190, 65, -84)$ pc for the PTS centre, consistent with the 3D dust mapping by \citet{Bialy+2021ApJL}. Throughout this work, the PTS centre serves exclusively as a spatial reference origin for computing $d_{\mathrm{PTS}}$ and defining the radial direction. All velocities were measured in the LSR frame, as defined in Sect.~\ref{YSOs selection and preprocessing}. We used fixed-width sliding windows with a width of 40 pc, a step of 6 pc, and a minimum $N > 19$. Within each sliding window, we formed an ensemble posterior probability density function by stacking the individual stellar posteriors with distance-dependent Gaussian weights. We report the median age and the 16th to 84th percentile range, with uncertainties estimated by bootstrap resampling. This sliding-window analysis provides an illustrative summary of the typical age as a function of $d_{\mathrm{PTS}}$. We quantified the slope using $d_{\mathrm{PTS}}$ and $t_{\mathrm{med}}$.

While systematic uncertainties in evolutionary models and extinction corrections inevitably affect the absolute age scale, our primary objective is not to establish definitive ages for discrete subgroups but rather to evaluate the macroscopic age gradient of the entire Perseus star-forming region relative to the PTS centre. This approach differs from \citet{Olivares+2023AandA}, which assigned ages collectively to seven kinematic groups via visual isochrone comparison in the faint regime ($G > 8$~mag). Our Bayesian framework operates at the individual star level to derive an age distribution for each source. This calculation explicitly accounts for both the per-star observational uncertainties ($\sigma_\mathrm{G}$, $\sigma_\mathrm{C}$, $\sigma_\mathrm{d}$, and $\sigma_{A_V}$) and a systematic model dispersion ($\sigma_\mathrm{m}$). By applying non-parametric statistics to these individual age estimates, our approach reliably traces the relative radial trend consistent with an inside-out progression, avoiding the spatial averaging inherent in group-level assignments.

\subsection{Kinematic diagnostics}
\label{Kinematic diagnostics}

We transformed the 3D ICRS Cartesian velocities into the LSR frame to get LSR 3D velocities $\boldsymbol{v}_{\mathrm{LSR}}$. To test the expansion-driven scenario, we constructed three complementary kinematic diagnostics. First, we probed whether individual stellar motions point systematically outwards from the PTS centre by measuring the angle between each star's $\boldsymbol{v}_{\mathrm{LSR}}$ and its radial vector. Then, we measured the intrinsic dispersion of the velocity component tangential to the plane of the sky and tested whether it declines with radius, as expected if the outer population is younger and has lower velocity dispersion. Finally, we verified these trends in full 3D velocity space, free from projection effects, to additionally reveal the radial dependence of the mean $\boldsymbol{v}_{\mathrm{LSR}}$ and the intrinsic 3D velocity dispersion.

\begin{figure}
\centering
\includegraphics[width=\columnwidth]{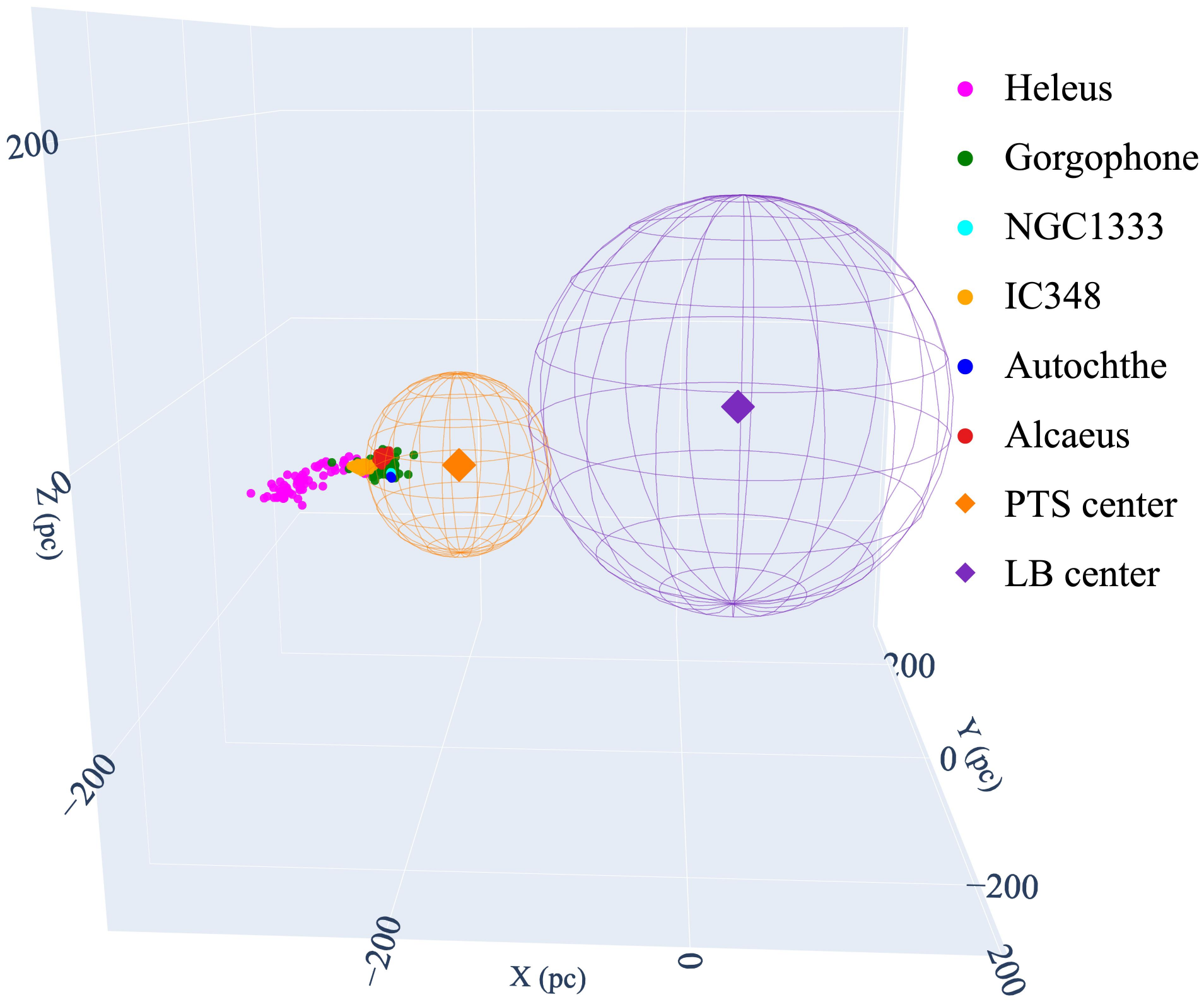}
\caption{3D spatial distribution of the selected Perseus YSOs, shown alongside the geometries of the PTS and the Local Bubble. This 3D interactive diagram is available online.} 
\label{plot}
\end{figure}

To diagnose the expansion kinematics, we defined the velocity-radial angle, $\theta$, for each star as the angle between its LSR 3D velocity vector and the radial vector pointing from the PTS centre to the star's position. Operating entirely in 3D space, this diagnostic is independent of line-of-sight projection effects. Then, we computed the tangential velocities by projecting these $\boldsymbol{v}_{\mathrm{LSR}}$ onto the local sky basis vectors in right ascension (${\rm RA}^\ast$) and declination (Dec), generating $v_{{\rm RA}^\ast}$ and $v_{\mathrm{Dec}}$ in $\mathrm{km\,s^{-1}}$, where ${\rm RA}^\ast$ denotes the conventional right-ascension tangential component on the celestial sphere \citep{Squicciarini+2021}. By evaluating kinematics using these linear tangential velocities, we removed the distance-induced $1/d$ scaling inherent in proper motions, enabling a direct evaluation of intrinsic stellar motions.  The same sliding window approach was used for the velocity dispersion analysis. To estimate the intrinsic tangential velocity dispersion $\sigma_{\mathrm{int}}$ within each sliding window, we accounted for the per-star measurement errors $e_i$. The observed sample dispersion $\sigma_{\mathrm{obs}}$ is a combination of the true intrinsic dispersion and this measurement noise. We estimated $\sigma_{\mathrm{int}}$ for the $v_{{\rm RA}^\ast}$ and $v_{\mathrm{Dec}}$ components separately using a heteroskedastic maximum likelihood estimator (MLE).

\begin{figure*}
\centering
\includegraphics[width=\textwidth]{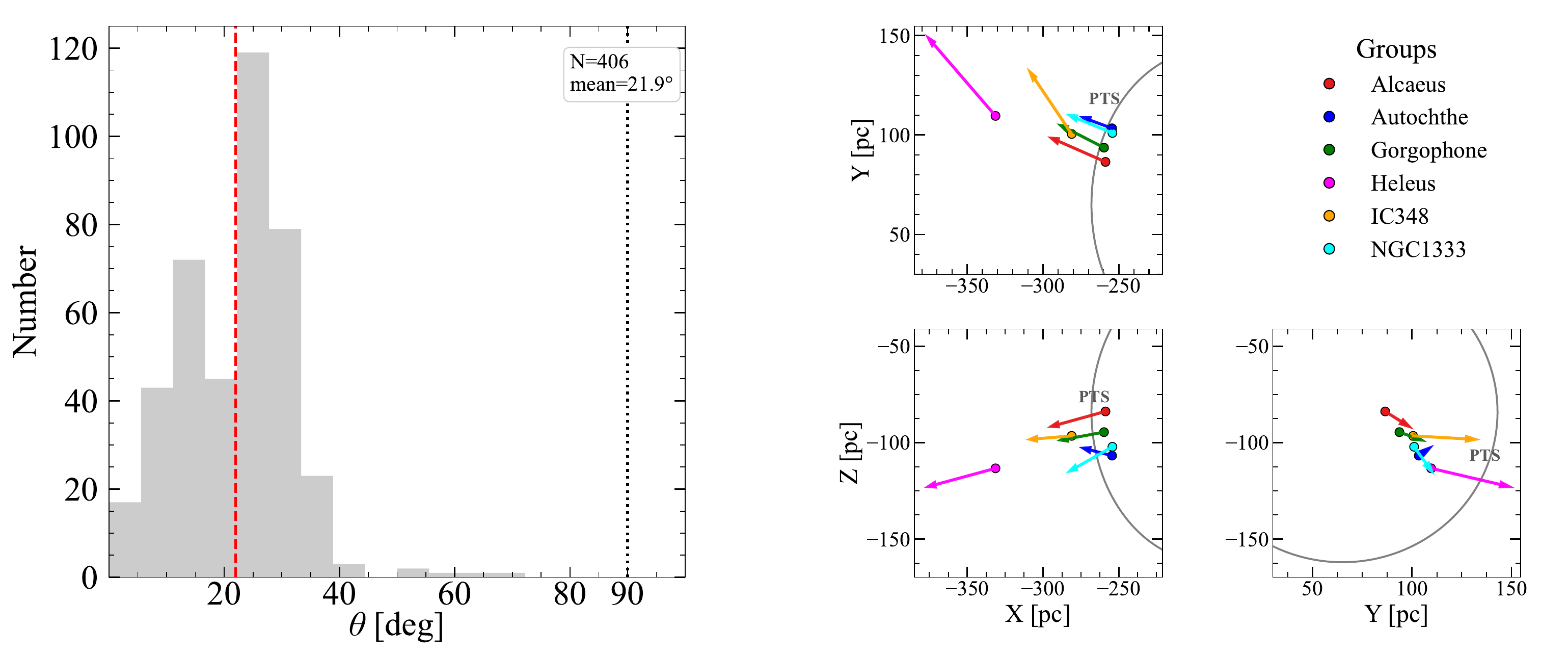}
\caption{Kinematic diagnostics of Perseus YSOs relative to the PTS centre. Left: Distribution of angles between each star's $\boldsymbol{v}_{\mathrm{LSR}}$ and the radial direction from the shell centre, with most stars moving outwards and no inward-pointing sources. The vertical dotted line at $90^{\circ}$ marks the threshold between outward and inward motion. The dashed red line represents the average angle of velocity.
Right: Mean $\boldsymbol{v}_{\mathrm{LSR}}$ for all kinematic groups in Galactic Cartesian coordinates. The grey arc labelled `PTS' marks the projected PTS boundary.}
\label{fig:vel-and-angle-results}
\end{figure*}

We employed a heteroscedastic model to address the distance-dependent structure of per-star velocity errors, where larger distances amplify proper-motion uncertainties. In our sample, $e_i$ therefore increases systematically with $d_{\mathrm{PTS}}$. Figure~\ref{fig:error_vs_distance} demonstrates this trend that the sliding median of $e_i$ rises from $\approx 0.55\;\mathrm{km\,s^{-1}}$ at $d_{\mathrm{PTS}} < 80\;\mathrm{pc}$ to $\approx 1.17\;\mathrm{km\,s^{-1}}$ at $d_{\mathrm{PTS}} > 140\;\mathrm{pc}$, a factor of $\sim\!2$ increase. Ignoring these per-star differences and assuming a single common error for all sources would systematically overestimate $\sigma_{\mathrm{int}}$ in the outer bins, where the larger $e_i$ values contribute a greater fraction of the observed dispersion. The heteroscedastic MLE described below avoids this bias by weighting each star according to its individual $e_i$.

This method assumes that the true velocities in a bin are drawn from a single Gaussian distribution $N(\mu, \sigma_{\mathrm{int}}^2)$, where $\mu$ represents the mean velocity. Each observed velocity $v_i^{\mathrm{obs}}$ is the sum of its true velocity and a measurement error $\epsilon_i$ drawn from $N(0, e_i^2)$. The likelihood of an observed star $i$ is therefore based on the model \(v_i^{\mathrm{obs}} \sim N(\mu, \sigma_{\mathrm{int}}^2 + e_i^2)\). The $e_i$ were derived by propagating the $\boldsymbol{v}_{\mathrm{LSR}}$ uncertainties via Monte Carlo simulation. We define the inverse-variance weights
\begin{equation}
w_i(\sigma_{\mathrm{int}}^2) = \frac{1}{\sigma_{\mathrm{int}}^2 + e_i^2} \,,
\end{equation}
which enters the 1D Gaussian likelihood for the observed
velocities in each bin. The corresponding log-likelihood for stars in each bin is
\begin{multline}
  \ln \mathcal{L}(\mu, \sigma_{\mathrm{int}})
  = -\frac{1}{2}\sum_{i}
  \Bigg[
    \ln\left[2\pi\left(\sigma_{\mathrm{int}}^2 + e_i^2\right)\right] \\
    +\left(v_i^{\mathrm{obs}} - \mu\right)^2
      \, w_i(\sigma_{\mathrm{int}}^2)
  \Bigg] \,.
\end{multline}
We maximized this log-likelihood function to obtain the optimal estimators for the $\mu$ and the intrinsic dispersion $\sigma_\mathrm{int}$. The final intrinsic tangential dispersion for the bin $\sigma_t$ is the quadrature combination of the two fitted components $\sigma_t = \sqrt{\sigma_{\mathrm{int, {\rm RA}^\ast}}^2 + \sigma_{\mathrm{int, Dec}}^2}$. 

This MLE framework is critical for distinguishing intrinsic kinematics from observational noise. By rigorously accounting for heteroscedastic errors, we uncover a previously unresolved radial decline in velocity dispersion, providing a new dimension of kinematic evidence for the shell's expansion that would otherwise remain hidden.

We further extended the same sliding-window and heteroscedastic MLE framework to the full 3D velocity field. Within each distance bin, we first subtracted the error-weighted mean velocity from each star's $\boldsymbol{v}_{\mathrm{LSR}}$ to obtain residual velocities. The intrinsic 3D velocity dispersion $\sigma_{\mathrm{3D}}$ was then estimated from these residuals using the same MLE procedure. In parallel, we computed the error-weighted mean $\boldsymbol{v}_{\mathrm{LSR}}$ as a function of $d_{\mathrm{PTS}}$ to characterize the radial velocity gradient.

To quantify radial trends throughout this work, we adopted the Theil-Sen robust estimator, defined as $\widehat{\beta}_{\mathrm{TS}} = \operatorname{median}_{\,i<j}\,\frac{y_j - y_i}{x_j - x_i}$, which represents the median of all pairwise slopes. Unlike ordinary least squares, whose slope estimate can be dominated by a few extreme points, the Theil-Sen median-based slope is inherently resistant to outliers, making it well-suited to astrophysical datasets where non-Gaussian tails arise from unresolved binaries, photometric variability, or small-number statistics in individual distance bins.

\subsection{Molecular gas kinematics from CO observations and astrometric alignment}
\label{sec:gas_data_method}

To investigate the kinematic relationship between the YSOs and the Perseus molecular cloud, we used $^{12}$CO ($J=1-0$) and $^{13}$CO ($J=1-0$) spectral line data cubes from the CO Coordinated Molecular Probe Line Extinction Thermal Emission Survey (COMPLETE) \citep{Ridge+2006AJ}. The data have an angular resolution of $\sim 46\arcsec$ and a spectral resolution of $\sim 0.06 \rm\,km\,s^{-1}$.

We generated the moment 1 map of the $^{12}$CO emission to trace the line-of-sight velocity field of the diffuse gas. Additionally, we computed the moment 0 map of the optically thinner $^{13}$CO emission to delineate the dense filamentary structures of the cloud. To accurately compare the spatial distribution of the YSOs with the molecular gas, it is necessary to align their coordinate epochs. The original Gaia DR3 astrometric parameters of the YSOs have a reference epoch of J2016.5. We propagated the celestial coordinates of the YSOs from J2016.5 backwards to J2000, which matches the observation epoch of the COMPLETE survey.

\begin{figure*}[ht!]
\centering
\includegraphics[width=\textwidth]{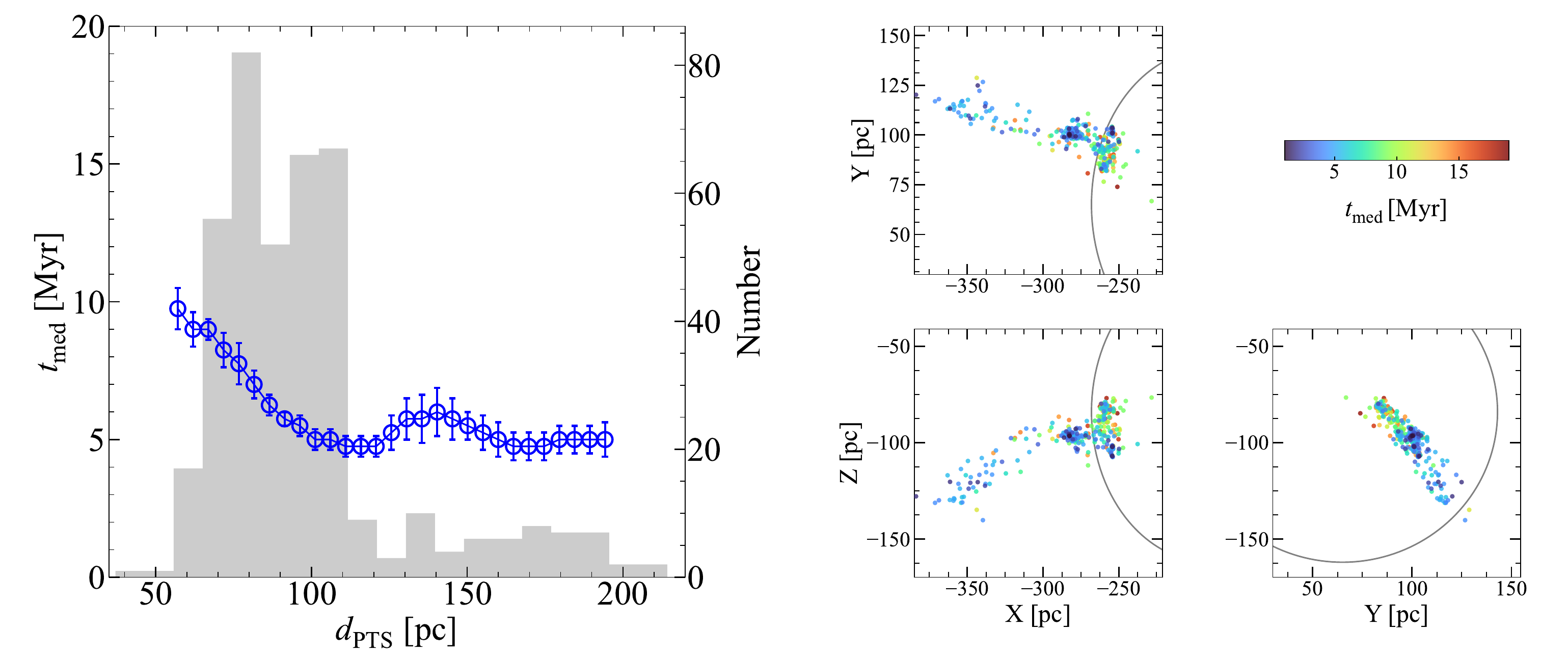}
\caption{Age-distance diagnostics for Perseus YSOs relative to the PTS centre. Left: Left y-axis showing sliding-window median ages $t_{\rm med}$ in megayears as a function of $d_{\mathrm{PTS}}$ from the shell centre. The points with the vertical error bars give an approximate one-sigma confidence range, while the right y-axis displays the number of YSOs in the background histogram. Right: Orthogonal projections of individual stars in Galactic Cartesian coordinates, colour-coded by the $t_{\mathrm{med}}$. The grey circle marks the projected PTS shell radius.}
\label{fig:combined-age-distance-analysis}
\end{figure*}

\begin{figure}
\centering
\includegraphics[width=\hsize]{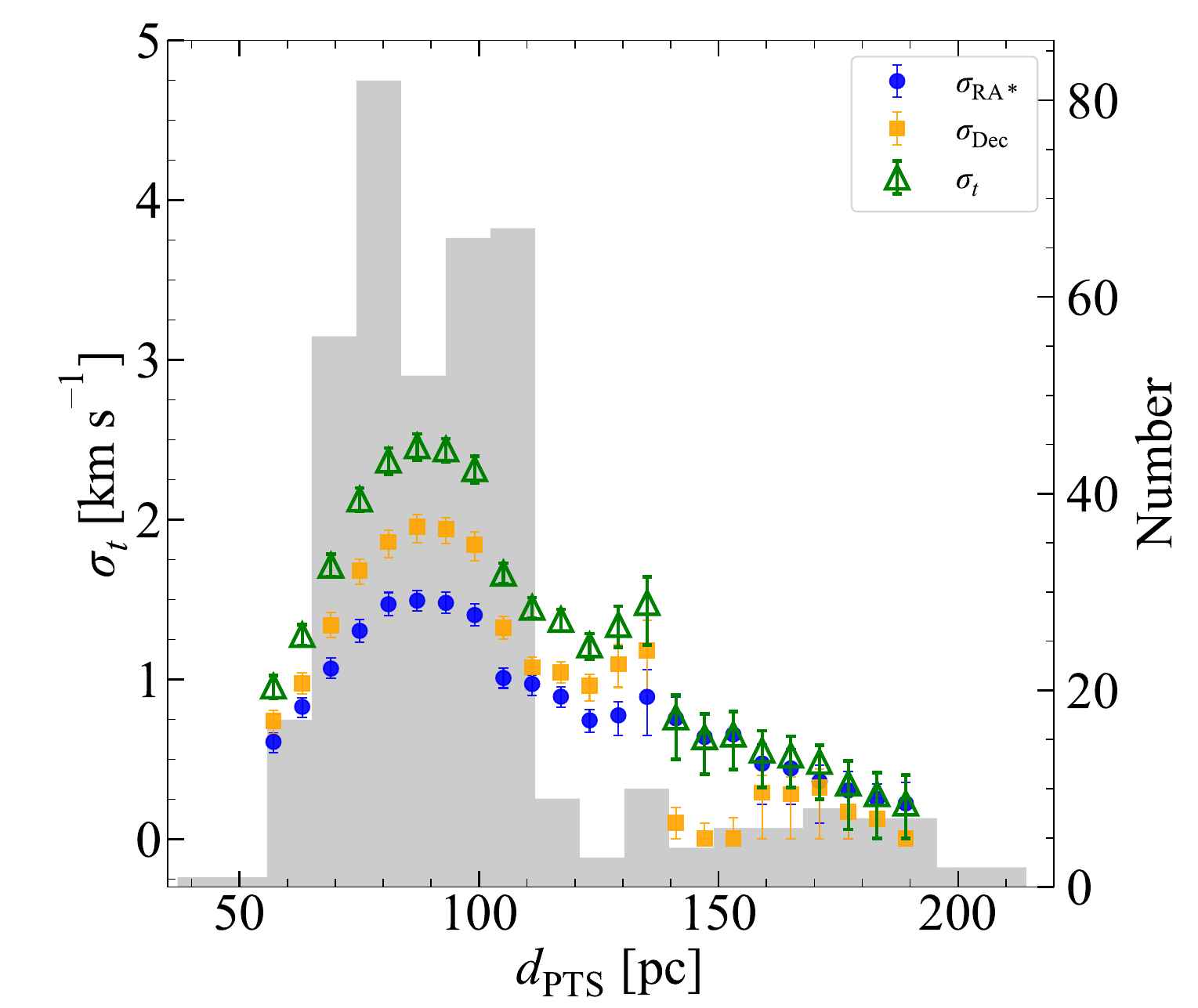}
\caption{Tangential velocity dispersion diagnostics for Perseus YSOs relative to the PTS centre. The blue circles and orange squares show the intrinsic velocity dispersions, $\sigma_{{\rm RA}^\ast}$ and $\sigma_{\rm Dec}$, of the two tangential components, while the green triangles show their quadrature combination, $\sigma_t$, as a function of the $d_{\rm PTS}$. The vertical error bars indicate the approximate one-sigma uncertainties.}
\label{fig:combined-kinematic-trends}
\end{figure}

\begin{figure*}[ht!]
\centering
\includegraphics[width=\textwidth]{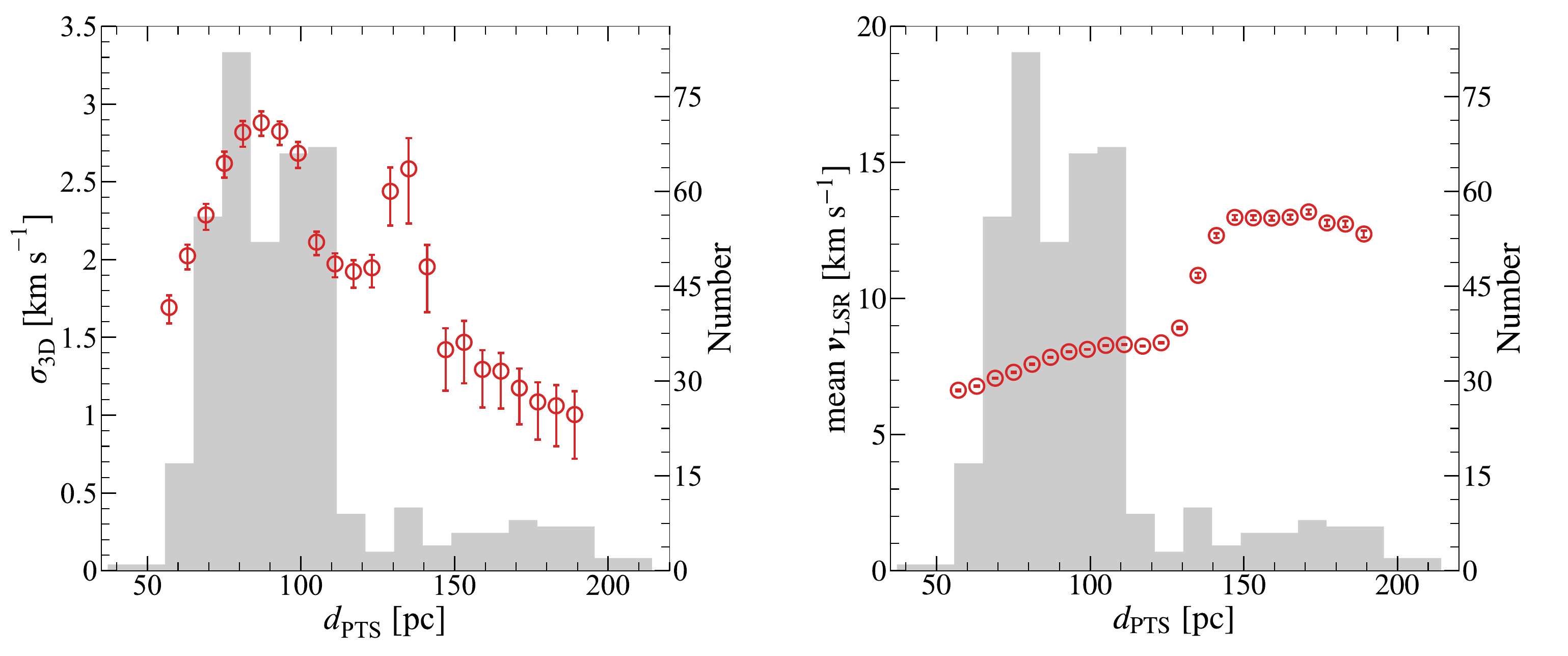}
\caption{3D kinematic diagnostics of the Perseus YSOs relative to the PTS centre. Left: Intrinsic 3D velocity dispersion as a function of $d_{\mathrm{PTS}}$, with the points and vertical error bars indicating the estimates and their approximate one-sigma uncertainties and a background histogram showing the number of sources per distance bin. Right: Error-weighted mean $\boldsymbol{v}_{\mathrm{LSR}}$ as a function of $d_{\mathrm{PTS}}$ with the same binning, one-sigma error bars, and background histogram.}
\label{fig:combined_3d_velocity_trends}
\end{figure*}

\section{Results}
\label{results}

In this section, we present the kinematic and temporal properties of the Perseus YSO population to reconstruct its star formation history. Based on the uniform 6D phase-space framework, we first evaluate the 3D spatial and velocity distributions, revealing a coherent outward motion of the YSOs relative to the PTS centre. We then discuss quantifying the radial trends in stellar ages and tangential velocity dispersions. To corroborate these 3D findings, we provide an independent projected diagnostic using infrared classifications. Finally, we compare the stellar proper motions with the CO gas velocity field to assess their large-scale kinematic coherence. Collectively, these multi-dimensional results establish the observational foundation for the expansion-driven scenario that will be discussed in Sect.~\ref{discussion}.

\subsection{3D Spatial distribution and kinematic trends}

To evaluate the influence of the PTS, we first established the physical geometry of the Perseus YSOs in 3D space relative to the shell's expansion centre. Figure~\ref{plot} presents the 3D spatial distribution of our 406 selected YSOs, alongside the models of the PTS and the Local Bubble. This 3D visualization reveals the complex spatial arrangement of the Perseus sub-regions along the line of sight. For example, based on recent high-precision Gaia DR3 surveys, groups such as Alcaeus are located at a closer heliocentric distance ($\sim 286$ pc) compared to the famous IC 348 ($\sim 315$ pc) and NGC 1333 ($\sim 292$ pc) \citep{Olivares+2023AandA}. Given this complex line-of-sight geometry, evaluating the stellar distribution and kinematics relative to the $d_{\mathrm{PTS}}$ provides the necessary physical framework to explore the expansion-driven scenario.

Figure~\ref{fig:vel-and-angle-results} presents the velocity-radial angle analysis. The left panel shows the distribution of $\theta$ for the YSO population, confirming a systematic expansion pattern that the entire population has $\boldsymbol{v}_{\mathrm{LSR}}$ pointing broadly away from the PTS centre, with all angles strictly $< 90^{\circ}$. The distribution peaks at small separation angles around 20–30$^{\circ}$, with the mean value of approximately 22$^{\circ}$. The right panel shows the mean $\boldsymbol{v}_{\mathrm{LSR}}$ vectors for each kinematic group in Galactic Cartesian coordinates, visually confirming the outward-pointing motions across the main substructures. 

\subsection{Age-distance relation}
\label{Age-distance relation}

In the left panel of Fig.~\ref{fig:combined-age-distance-analysis}, we show the correlation between the median stellar age, $t_{\mathrm{med}}$, and the 3D distance to the PTS centre, $d_{\mathrm{PTS}}$, for the 406 YSOs. The sliding-window analysis described in Sect.~\ref{data_methods} reveals a clear radial pattern that $t_{\mathrm{med}}$ declines nearly monotonically with increasing $d_{\mathrm{PTS}}$, indicating progressively younger populations at larger radii. The right panel shows orthogonal projections of the same 406 YSOs in Galactic Cartesian coordinates, colour-coded by $t_{\mathrm{med}}$. This spatial view demonstrates that the age gradient is not driven by a few outliers but reflects a coherent large-scale structure across the PTS region. Then, we performed an independent robustness check using the 406 individual stars. We tested the trend between individual $t_{\mathrm{med}}$ and $d_{\mathrm{PTS}}$ using distribution-free estimators, which make no Gaussian or linearity assumptions. The slope was estimated with the robust Theil-Sen estimator (Sect.~\ref{Kinematic diagnostics}). We further quantified the monotonic association using rank statistics with standard large-sample $p$-value approximations. Additional robustness checks based on resampling are presented in Fig.~\ref{fig:robust_slope_bootstrap} of Appendix~\ref{Robustness}. The age-distance slope also remains negative when accounting for plausible uncertainties in the 3D position of the PTS centre (Fig.~\ref{fig_jointage} of Appendix~\ref{app_robustness_joint}).

To quantify the age-distance trend using individual stars, we adopted the robust Theil-Sen estimator. The age-distance trend is strongly negative, and we find that the median-slope estimate is \(\frac{\mathrm{d}\,t_{\mathrm{med}}}{\mathrm{d}\,d_{\mathrm{PTS}}}
= -0.05~\mathrm{Myr\,pc^{-1}}\) with a $95\%$ confidence interval $\big[-0.07,\,-0.04\big]~\mathrm{Myr\,pc^{-1}}$ entirely below zero. We evaluated YSO ages as a continuous function of distance instead of separating them into discrete kinematic groups \citep{Olivares+2023AandA}. This approach reveals a continuous radial age gradient that physically unifies the stellar populations. The results demonstrate that stars located at larger radii are progressively younger. This observed inside-out chronological sequence is quantitatively consistent with the predictions of an expansion-driven triggering model. The inverse of this slope directly yields a characteristic velocity scale $1/|\widehat{\beta}_{\mathrm{TS}}| \approx 20~\mathrm{pc\,Myr^{-1}}$. Applying the standard unit conversion ($1~\mathrm{pc\,Myr^{-1}} \approx 0.978~\mathrm{km\,s^{-1}}$), we obtain a kinematic propagation speed of $v_{\mathrm{TS}} \approx 20~\mathrm{km\,s^{-1}}$, with a 95\% confidence interval of approximately $14$--$25~\mathrm{km\,s^{-1}}$. The physical interpretation of this velocity scale is discussed in Sect.~\ref{Time}.

\begin{figure*}[ht!]
\centering 
\includegraphics[width=\textwidth]{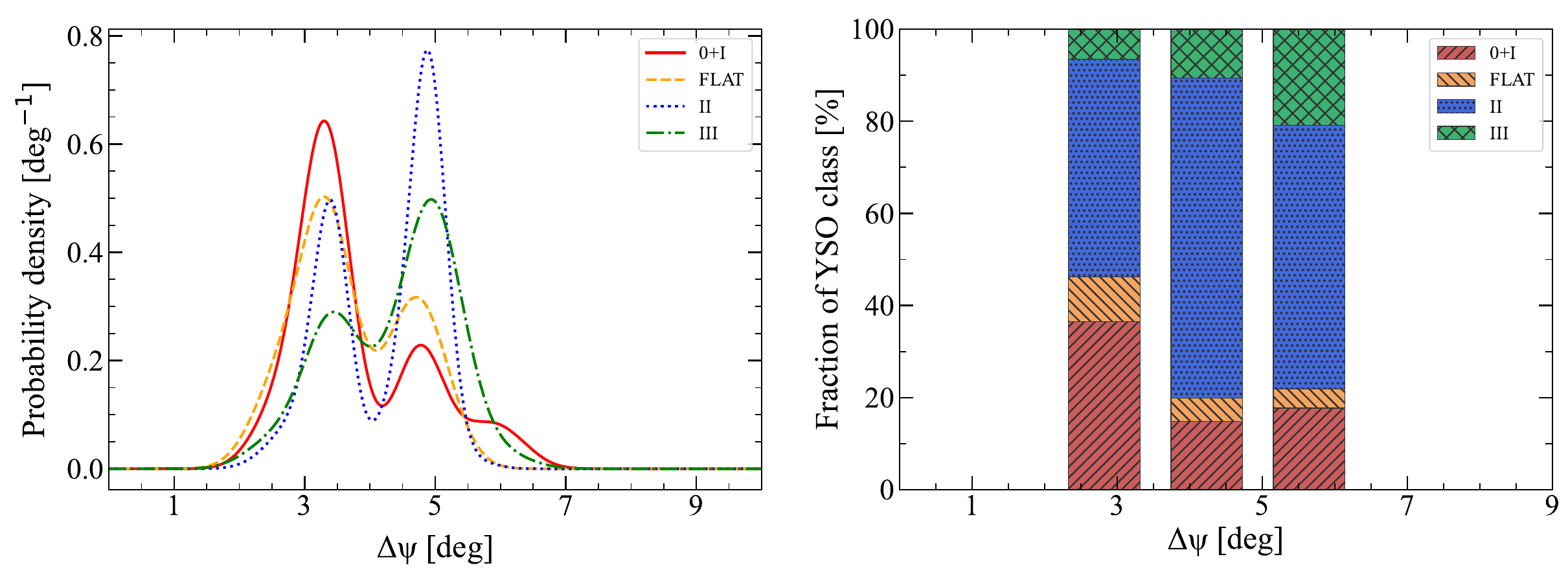}
\caption{Projected angular diagnostics for the Perseus YSOs relative to the PTS centre. Left: Distribution of separation angles between the stellar positions and the PTS centre on the sky, with histograms and smooth probability density curves for different YSO classes. Right: Class composition as a function of separation angle, with the fractions of the 0+I, FLAT, II, and III classes plotted in angle bins across the full range of offsets.}
\label{fig:combined-projected-trends}
\end{figure*}

\subsection{Velocity dispersion-distance relation}
 \label{Tangential-velocity dispersion}

We employed the sliding window MLE analysis detailed in Sect.~\ref{data_methods} to calculate the intrinsic tangential velocity dispersion $\sigma_t$ by fitting the dispersion of the components $v_{{\rm RA}^\ast}$ and $v_{\mathrm{Dec}}$ in each bin. The resulting radial trend of $\sigma_t$ is presented in Fig.~\ref{fig:combined-kinematic-trends}. The figure shows that $\sigma_t$ decreases with increasing $d_{\mathrm{PTS}}$, revealing a corresponding kinematic radial pattern. Additional robustness checks based on resampling are presented in Fig.~\ref{fig:trend_sigma_t_bootstrap} of Appendix~\ref{Robustness}. The $\sigma_t$--$d_{\rm PTS}$ slope remains negative when accounting for plausible uncertainties in the 3D position of the PTS centre (Fig.~\ref{jointvdisp} of Appendix~\ref{app_robustness_joint}).

Quantitatively, the kinematic trend is strongly negative. The robust Theil-Sen slope estimate is
\(\frac{d\,\sigma_t}{d\,d_{\rm PTS}} \;=\; -0.016~{\rm km\,s^{-1}\,pc^{-1}}\),
with a confidence interval of 95\% of \(\big[-0.02,\,-0.011\big]~{\rm km\,s^{-1}\,pc^{-1}}\) lying entirely below zero.
Bootstrap tests confirm that this radial decline in \(\sigma_t\) is not driven by a few outliers, indicating that the outer population has lower velocity dispersion. This kinematic behaviour provides independent dynamical evidence for the inside-out sequence. The higher dispersion of the inner population and lower dispersion of the outer generation align directly with the kinematic predictions of an expansion-driven star formation scenario triggered by the PTS.

To verify that these trends are not artefacts of sky-plane projection, we extended the analysis to the full 3D velocity field using the framework described in Sect.~\ref{Kinematic diagnostics}. Figure~\ref{fig:combined_3d_velocity_trends} presents the intrinsic 3D velocity dispersion and the error-weighted mean $\boldsymbol{v}_{\mathrm{LSR}}$ as a function of $d_{\mathrm{PTS}}$. The $\sigma_{\mathrm{3D}}$ follows essentially the same radial trend as $\sigma_t$, with elevated values in the inner region and a general decline towards larger $d_{\mathrm{PTS}}$. Quantitatively, the robust Theil-Sen slope estimate is $\frac{d\,\sigma_{\mathrm{3D}}}{d\,d_{\mathrm{PTS}}} = -0.013~\mathrm{km\,s^{-1}\,pc^{-1}}$, with a 95\% confidence interval of $[-0.016,\,-0.008]~\mathrm{km\,s^{-1}\,pc^{-1}}$ lying entirely below zero, confirming that the radial cooling signature is robust in 3D. Furthermore, the mean $\boldsymbol{v}_{\mathrm{LSR}}$ exhibits a monotonic positive correlation with $d_{\mathrm{PTS}}$, with a Theil-Sen slope of $+0.050~\mathrm{km\,s^{-1}\,pc^{-1}}$ and a 95\% confidence interval of $[+0.036,\,+0.064]~\mathrm{km\,s^{-1}\,pc^{-1}}$, indicating that the outer populations have systematically higher bulk velocities.

\subsection{Direction-only consistency with YSO catalogue}

To complement the age-distance and tangential velocity-dispersity trends, we performed an independent, purely projected test using the infrared YSO catalogue. We examined the angular separation \(\Delta\psi\) between each star and the PTS-centre direction,
\begin{equation}
\cos\Delta\psi \;=\; \sin\delta\,\sin\delta_c \;+\; \cos\delta\,\cos\delta_c\,\cos(\alpha-\alpha_c) \,,
\end{equation}
where \((\alpha_c,\delta_c)\) denote the ICRS RA and declination of the PTS centre. We constructed density normalized histograms with Gaussian Kernel Density Estimation (KDE) curves for each YSO class over a small angle domain. We also defined three concentric annuli around the reference position. The boundaries were chosen so that the annuli have equal width in projected angular separation $\Delta\psi$ and together cover the entire range of the sample.

Figure~\ref{fig:combined-projected-trends} illustrates the projected angular structure. In the left panel, the KDE profiles reveal that the younger classes (combined Class 0 and I and flat) peak at smaller separations ($\Delta\psi \approx 3.4^\circ$), while the older classes (Class II and III) peak at larger separations ($\Delta\psi \approx 4.6^\circ$), with Class II reaching the highest absolute density owing to its largest sample size. The evolutionary trend is further quantified in the right panel, where the class composition plot reveals a clear gradient: the relative fraction of the combined Class 0 and I
and flat sources is highest in the innermost annulus and decreases at larger separations, whereas Class III shows the opposite behaviour. These diagnostics confirm that younger YSOs are relatively more concentrated towards the PTS-centre direction, providing a purely projected consistency check for the inside-out sequence.

These diagnostics confirm that younger YSOs are tightly concentrated in the PTS centre direction. The older sources present a highly dispersed configuration. This 2D finding independently supports our primary 3D results. Provides a purely projected consistency check for the negative radial gradients in age and tangential velocity dispersion. The projected spatial segregation of different evolutionary stages is compatible with an expansion-driven outward sequence of star formation on the PTS.

\begin{figure*}[ht!]
\centering 
\includegraphics[width=\textwidth]{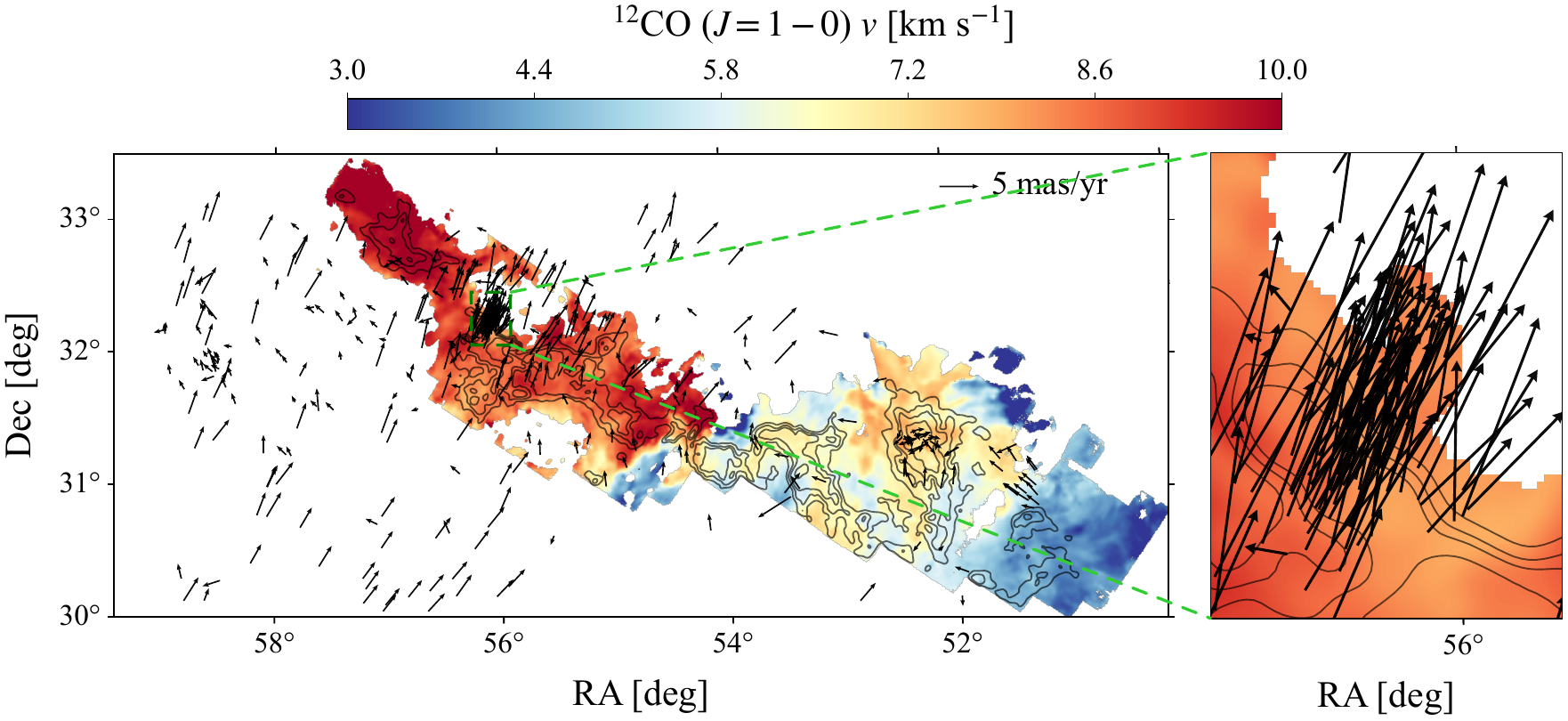}
\caption{Composite kinematic view of the Perseus molecular cloud. Left: $^{12}$CO ($J=1-0$) moment-1 velocity field ($v$) from the COMPLETE survey, overlaid with $^{13}$CO ($J=1-0$) moment-0 integrated-intensity contours tracing the dense-gas structure. The black arrows indicate the subtracted solar peculiar motion vectors on the plane of sky and a reference vector of $5\,\mathrm{mas\,yr^{-1}}$ (upper right). The dashed green rectangle marks the zoom-in region of IC~348 shown in the right panel. This zoom-in region is shown to better visualize the IC~348 region, as it contains the highest source density in our sample (160 YSOs); otherwise, the proper-motion vectors would highly overlap.}
\label{fig:zoom}
\end{figure*}

\begin{figure}
\centering
\includegraphics[width=\hsize]{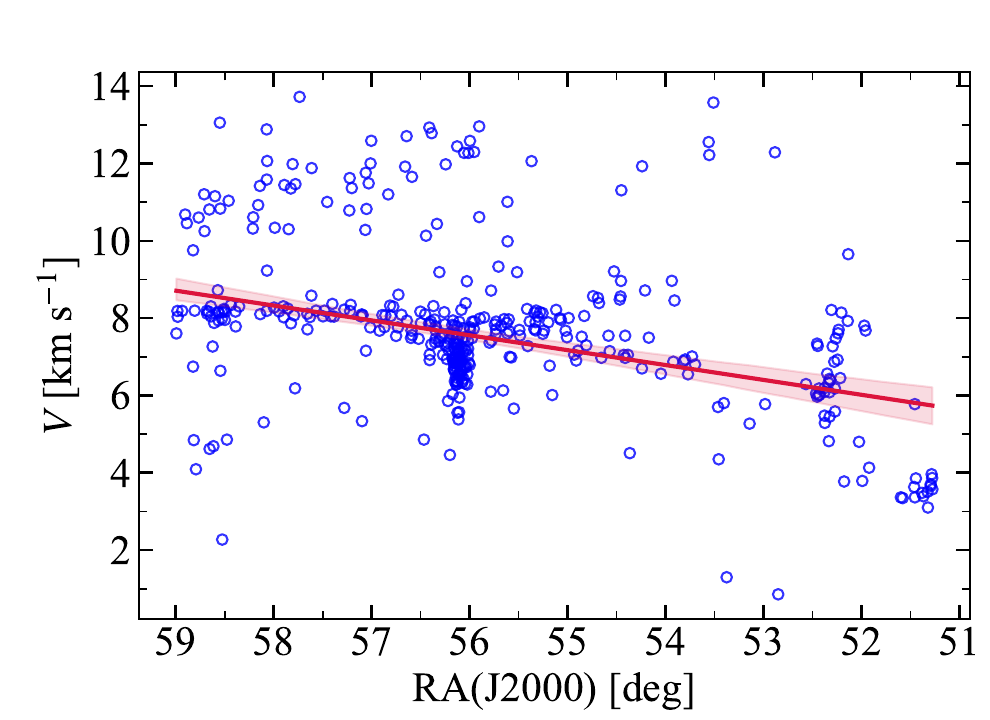}
\caption{Line-of-sight velocity changes along the RA.
The open circles show the LSR-corrected line-of-sight velocities of individual YSOs ($V$) as a function of ${\rm RA}({\rm J2000})$. The solid red line shows the robust Theil-Sen linear fit, and the shaded area indicates the 95\% bootstrap confidence interval. The systematic increase in $V$ towards larger RA is consistent with the large-scale $^{12}$CO $V_{\rm LSR}$ gradient seen in the moment-1 map.}
\label{YSO_vlos_LSR_vs_RA}
\end{figure}

\subsection{Kinematic coherence between YSOs and molecular gas}
\label{kinematic_coupling}

We compared the kinematics of the YSOs with the velocity field of the Perseus molecular cloud to further investigate the dynamical relationship between the young stellar populations and their natal environment. Figure~\ref{fig:zoom} presents a composite view of the region. The map overlays the Gaia proper-motion vectors on the $^{12}$CO moment-1 map of the LSR radial velocity $V_{\rm LSR}$ and the $^{13}$CO moment-0 integrated-intensity contours from the COMPLETE survey. We propagated the YSO positions to the CO map epoch J2000.0 using the Gaia proper motions to ensure spatial registration with the gas emission.

Spatially, the high-column-density $^{13}$CO contours trace the dense filamentary skeleton of the cloud, coinciding with the primary concentrations of YSOs (e.g. NGC~1333 and IC~348). A fraction of the distributed YSO population lies outside the dense gas structures. Kinematically, the YSO proper motions exhibit a coherent macroscopic bulk flow across the sky plane rather than a radially dispersive pattern. In the background, the $^{12}$CO gas exhibits a continuous radial velocity gradient along the primary axis of the cloud.

To quantify the kinematic coupling in 3D velocity field, we examined the line-of-sight velocities. Figure~\ref{YSO_vlos_LSR_vs_RA} shows the LSR-corrected line-of-sight velocities of individual YSOs ($V$) as a function of RA. We quantified the gradient using the robust Theil-Sen estimator, consistent with the methodology applied in Sects.~\ref{Age-distance relation} and~\ref{Tangential-velocity dispersion}. The resulting slope is $0.39~\mathrm{km\,s^{-1}\,deg^{-1}}$, with a 95\% confidence interval of $[0.31,\,0.47]~\mathrm{km\,s^{-1}\,deg^{-1}}$ lying entirely above zero, confirming that the positive gradient is significant despite the intrinsic velocity dispersion of the YSO population. The shaded areas in the figure shows the 95\% bootstrap confidence band of the fit. We observe a systematic linear relation, indicating an increase of $V$ towards larger RA. Crucially, this stellar line-of-sight velocity gradient spatially and quantitatively matches the large-scale $^{12}$CO $V_{\rm LSR}$ gradient seen in the moment-1 map.

The morphological alignment of the 2D coherent bulk flow, combined with the line-of-sight velocity correlation between the stars and the gas, provides robust evidence for macroscopic kinematic coherence. The absence of a highly dispersive signature in the proper motions is consistent with previous high-precision astrometric studies that found no clear evidence of internal expansion within the individual subclusters of Perseus \citep{Ortiz-Leon+2018ApJ}. This shared 3D kinematic behaviour extends the localized core and gas coherence previously observed in Perseus \citep{Kirk+2010ApJ} to the more evolved distributed YSO population. Collectively, these coherent kinematics demonstrate that the entire Perseus complex shares a consistent flow, consistent with both the gas and the embedded stars being swept up and driven by the expansion of the PTS.

\section{Discussion}
\label{discussion}

The 6D phase-space data utilized in this study consist of a meticulously screened, high-quality sample, enabling a robust analysis of the kinematics and accurate age determination. Our results regarding the overall velocity field are consistent with previous YSO studies, yet provide new constraints on their temporal evolution. Based on the converging lines of evidence presented here, we conclude that the formation of YSOs in the Perseus region was likely triggered by the past expansion of the PTS.

\subsection{Sample completeness and spatial substructure}
\label{discussion_sample}

A primary limitation concerns the sample size and completeness. Stringent quality criteria restrict our sample to 406 YSOs, representing approximately 40 percent of the total Perseus census \citep{Olivares+2023AandA}. Furthermore, current 3D reconstructions of the PTS lack full sky coverage. Combined with the shell's apparently low expansion speed, dynamical signatures remain subtle and challenging to disentangle from measurement noise, projection effects, and the intrinsic selection incompleteness of faint sources. Integrating kinematics with spatial distribution and age analysis is essential to maximize the physical insights derived from this limited sample.

Although the exclusion of the extended Perseus groups Electryon, Mestor, and Cynurus is intrinsically determined by the spatial coverage of the parent catalogue (see Sect.~\ref{YSOs selection and preprocessing}), this limitation physically aligns with the scope of the expansion-driven scenario. The exclusion of Electryon, Mestor, and Cynurus is physically justified by their spatial and dynamical decoupling from the PTS. Physically, Electryon and Mestor are located far from the PTS and their formation is spatially decoupled from the shell's expansion \citep{Wang+2022ApJ}. Cynurus is similarly located on the distant periphery of the complex and is highly dispersed \citep{Kounkel+2022AJ}, placing it outside the main region of kinematic coherence. While \citet{Wang+2022ApJ} also categorized Heleus as distant from the shell, our 3D kinematic analysis shows that all Heleus members have velocity-radial angles $\theta < 90^\circ$, indicating coherent outward motions consistent with the PTS expansion pattern and warranting its retention. Excluding the physically decoupled and peripheral groups ensures that our sample remains kinematically coherent for testing the expansion-driven scenario.

We note that the Perseus molecular cloud is situated on the far side of the PTS, with the PTS residing directly between the Local Bubble and Perseus \citep{Bialy+2021ApJL}. Due to this macroscopic geometric configuration, we do not consider the direct kinematic influence of the Local Bubble on Perseus. While recent studies reveal that the star formation history in the Taurus region on the near side of the shell is predominantly driven by the Local Bubble expansion \citep{Liu+2025SCPMA}, the coherent $v \propto r$ expansion gradient and the age-distance relation we observe in Perseus align consistently with the specific 3D geometric centre of the PTS. This macroscopic geometric configuration, combined with the coherent radial expansion gradient, indicates that the kinematic signatures in Perseus are primarily governed by the local expansion of the PTS. In this specific local environment, the direct kinematic impact of the Local Bubble is likely a secondary effect rather than the primary driver.

\begin{figure}
\centering
\includegraphics[width=\columnwidth]{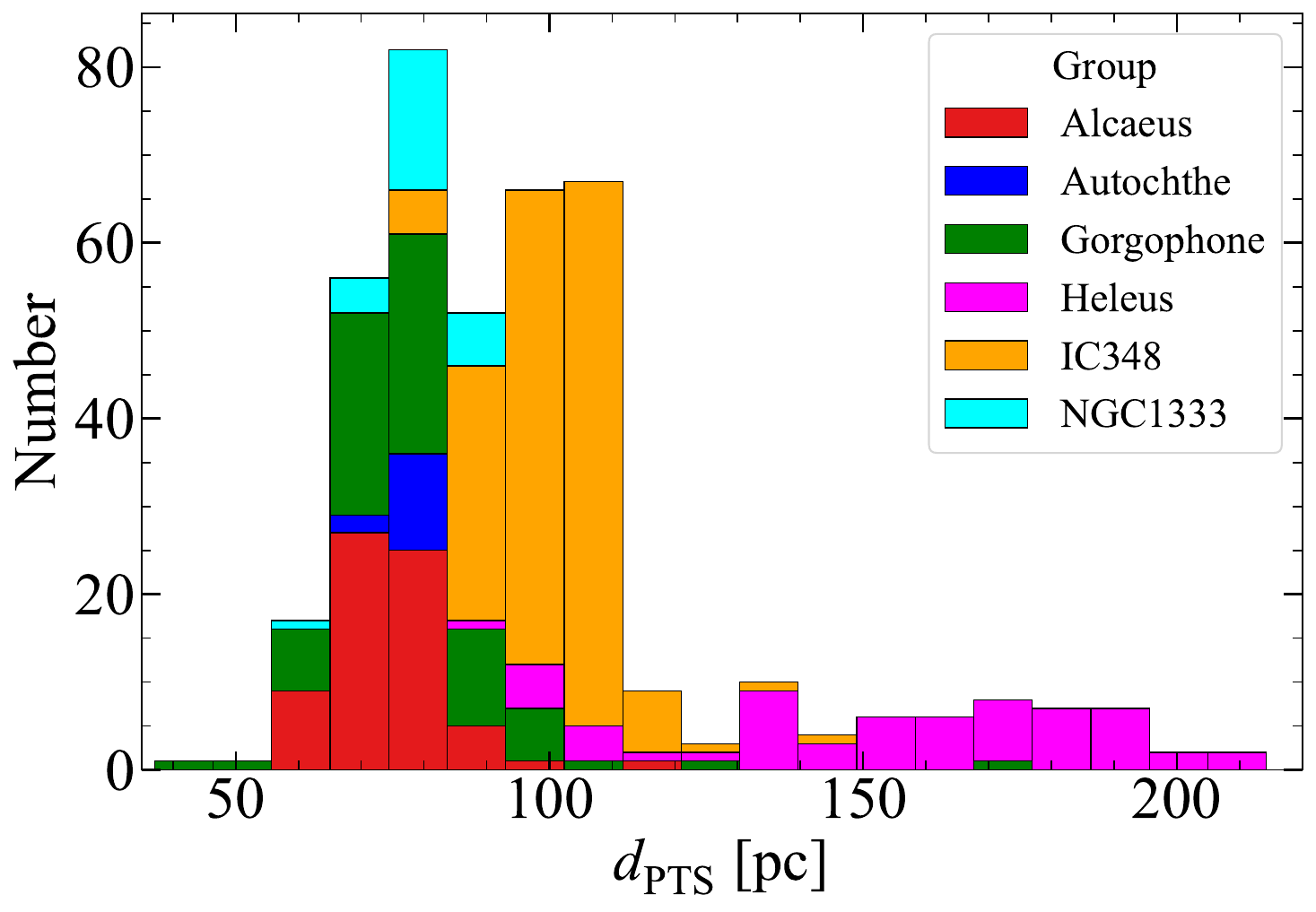}
\caption{Stacked histogram of the 3D distance ($d_{\rm PTS}$) to the PTS centre for all YSOs. The different colours represent the six kinematic groups.}
\label{distance_to_pts_histogram}
\end{figure}
\subsection{Timescales and triggering mechanics}
\label{Time}

Examining the spatial distribution decomposed by kinematic group reveals clear transitions in the stellar population composition along the $d_{\mathrm{PTS}}$. Specifically, we observe a distinct shift from the Alcaeus, Autochthe, Gorgophone, and NGC~1333 associations to IC~348 between $\sim$55~pc and $\sim$100~pc, followed by a transition from IC~348 to Heleus beyond $\sim$125~pc, as shown in Fig.~\ref{distance_to_pts_histogram}. This evident spatial segregation of kinematic groups supports our proposed scenario of layered star formation driven by the expansion of the PTS.

While the large-scale spatial sequence supports a global triggering scenario, the relative youth of NGC 1333 ($\sim 1-3$ Myr) \citep{Wilking+2004AJ,Wang+2022ApJ} compared to IC 348 ($\sim 3-6$ Myr) \citep{Bell+2013MNRAS,Wang+2022ApJ,Olivares+2023AandA} may appear to contradict a strictly unidirectional propagation along the $d_{\mathrm{PTS}}$ axis. However, by evaluating the $t_{\mathrm{med}}$ as a continuous function of spatial distance, our analysis captures the chronological sequence of the Perseus complex, rather than comparing the discrete histories of predefined kinematic groups. Physically, superbubble expansion into a clumpy molecular cloud is a non-linear process \citep{MacLow+2004RvMP,Krause+2020SSRv}. While the PTS provides a macroscopic global trigger for the Perseus complex, local density variations within the cloud can lead to delayed shock compression or localized, rapid secondary epochs of star formation \citep{Walch+2015MNRAS}. This localized rapid collapse explains the youth and high density of NGC 1333, reflecting complex gas dynamics rather than negating the overarching expansion-driven gradients imprinted by the PTS.

Several phenomena of the Perseus region remain under debate. One study analysed distributed YSOs in the Perseus molecular cloud using Gaia and LAMOST and reported a gradient-like signal in the spatial and/or kinematic distributions \citep{Wang+2022ApJ}. Another study reconstructed the dynamical star-forming history of Per OB2 and discussed candidate runaway stars, their membership, and likely origins \citep{Kounkel+2022AJ}. In parallel, current PTS shell models lack robust constraints on geometry and age, and plausible supernova triggers or progenitor clusters have not been conclusively identified.
Our analysis emphasizes shell-centric radial trends in ages and tangential kinematics, providing complementary, expansion-oriented constraints on the PTS system.

The observed outward velocity vectors align with the PTS centre, consistent with the geometric predictions of the expanding shell model proposed for the PTS \citep{Bialy+2021ApJL}, while the age–distance slope remains negative and $\sigma_t$ declines with radius across robustness checks. This joint temporal-kinematic pattern is difficult to reconcile with a localized collision or geometries tied to other large-scale structures. While alternative explanations cannot be completely excluded, the combined constraints from the geometry, age distribution, and kinematics are most naturally explained by a PTS-driven, inside-out sequence.

Interpreting the negative age-distance slope as an outwardly propagating triggering front along the PTS, the characteristic velocity $v_{\mathrm{TS}} \simeq 20~\mathrm{km\,s^{-1}}$ derived in Sect.~\ref{Age-distance relation} represents the propagation speed of the star-formation trigger during the shell's active expansion phase, rather than the current bulk motion of the stars or gas. Previous estimates suggest a broad age range for the PTS of $t_{\mathrm{age}} \approx 6\text{-}22$ Myr, based on the assumption that the shell has currently decelerated to the ambient turbulent velocity, $v_{\mathrm{tb}} \approx 7~\mathrm{km\,s^{-1}}$ \citep{Bialy+2021ApJL}. Our higher derived propagation speed is broadly consistent with this evolutionary scenario, effectively probing the earlier epoch when the shell was expanding significantly faster than $v_{\mathrm{tb}}$. For YSOs observed at distances of $d_{\mathrm{PTS}} \approx 60\text{-}200$\ pc, a propagation speed of $\sim 14\text{-}25~\mathrm{km\,s^{-1}}$ implies kinematic timescales of $\sim 3\text{-}14$\ Myr. This chronology is consistent with the observed inside-out age gradient-where YSOs at larger radii are younger ($\lesssim\!10$\ Myr) and sits comfortably within the lower bounds of the PTS age estimates derived from dust observations. Moreover, a comparable oscillatory morphology resembling the decrease-increase-decrease trend found in our age analysis has recently been reported for Taurus YSOs relative to the Local Bubble centre \citep{Liu+2025SCPMA}. Given that both Perseus and Taurus reside on the PTS surface, this shared feature suggests a common structural origin linked to the complex shell geometry.

In Sect.~\ref{Tangential-velocity dispersion}, we show that the tangential velocity dispersion, $\sigma_t$, decreases with $d_{\mathrm{PTS}}$ and that this pattern indicates larger velocity dispersion in the inner populations and smaller dispersion in the outer ones. A related question is whether these trends persist when we move from the projected tangential kinematics to the full 3D velocity field. Figure~\ref{fig:combined_3d_velocity_trends} presents the mean $\boldsymbol{v}_{\mathrm{LSR}}$ and the intrinsic 3D velocity dispersion as a function of $d_{\mathrm{PTS}}$, computed using the framework described in Sect.~~\ref{Kinematic diagnostics}. The 3D velocity dispersion follows essentially the same radial trend as $\sigma_t$, with elevated values in the inner region and a general decline towards larger $d_{\mathrm{PTS}}$. Furthermore, the mean $\boldsymbol{v}_{\mathrm{LSR}}$ exhibits a monotonic positive correlation with $d_{\rm PTS}$. While such velocity gradients ($v \propto r$) can arise from internal dynamical relaxation or longitudinal gravitational collapse in elongated clouds, the coherent, unidirectional bulk flow of Perseus disfavours a purely internal kinematic origin. Instead, within the framework of the PTS expansion, this linear increase physically indicates that the younger, outer YSOs near the active shock front retain higher drift velocities, whereas the older, inner populations have dynamically relaxed and kinematically decoupled from the macroscopic flow over their longer evolutionary timescales.

\begin{figure}
\centering
\includegraphics[width=\columnwidth]{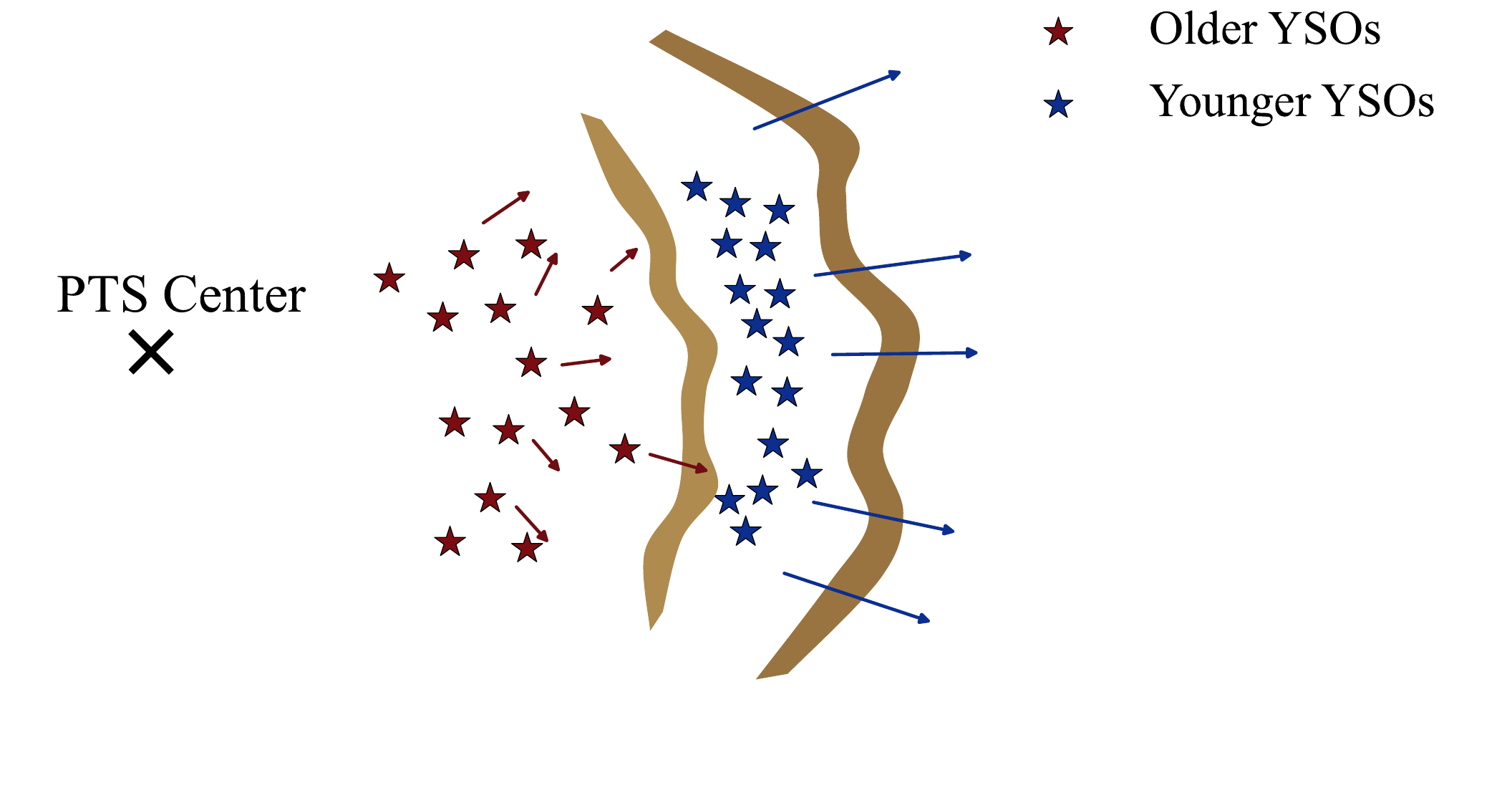}
\caption{Schematic of the expansion-driven star formation scenario proposed for the Perseus region. The PTS centre represents the origin of the historical shell expansion, with the outward direction indicating the propagation of the expanding shell. The younger YSO populations are located closer to the outer shell boundary, whereas older populations are distributed in the inner region, illustrating the proposed inside-out sequence of star formation.}
\label{expansion}
\end{figure}

In both tangential and 3D velocity dispersion profiles, distinct double peaks are observed at $\sim$90~pc and $\sim$135~pc (Figs.~\ref{fig:combined-kinematic-trends} and~\ref{fig:combined_3d_velocity_trends}). We find that these kinematic peaks coincide remarkably well with the spatial intervals separating distinct kinematic groups in the source density distribution. This correspondence indicates that the observed double-peak morphology is fundamentally driven by the spatial segregation of distinct stellar aggregates rather than being an artefact of sampling. Although the region beyond $d_{\mathrm{PTS}} \gtrsim 125$~pc marks a transition zone between different kinematic groups, we cannot exclude the possibility that the fluctuations there originate from stochastic noise reflecting the low stellar density. Collectively, these diagnostics reinforce the picture of an inner population with larger velocity dispersion and an outer generation with smaller velocity dispersion, consistent with a layered sequence of star formation triggered by the expanding PTS.

This expansion-driven star formation scenario is illustrated by the sketch in Fig.~\ref{expansion}. While both the Perseus and Taurus molecular clouds reside on the opposing sides of the PTS \citep{Bialy+2021ApJL}, our conceptual model depicts the physical processes primarily on the Perseus side. This underlying mechanism of expansion-driven star formation is deduced directly from the observed YSO kinematics and radial trends in Perseus. Although Taurus is also part of the PTS, it is not discussed in our proposed picture because recent studies have revealed that star formation in the Taurus region is predominantly driven by the expansion of the Local Bubble \citep{Liu+2025SCPMA}. Consequently, Fig.~\ref{expansion} simply visualizes our proposed physical scenario, illustrating how the outward shock front might trigger sequential star formation and shape the observed age and kinematic gradients.

Future work should go beyond pure kinematics to test the expansion-triggered picture. Dense gas kinematics and column density maps can quantify the relative roles of magnetic pressure, turbulence, and momentum injection in shaping the shell and potentially triggering star formation \citep{Chen+2024ApJ}. Placing the age-distance and dispersion-distance relations alongside radial trends in magnetic and gas structure will enable stronger, multi-physics tests of the outward-younger, less dispersed sequence and help assess causal links between the PTS and the Perseus star-forming complex. Recent advances in 3D magnetic-field studies have begun to reveal the cloud's complex geometry. The first 3D field reconstruction of Perseus shows a concave magnetic morphology that is consistent with the cloud being swept up by the PTS, while on cluster scales, polarization data in IC 348 indicate that magnetic fields  regulate filamentary fragmentation and gravitational collapse \citep{Tahani+2022A&A,Choi+2024ApJ}. Jointly modelling the Taurus and Perseus molecular clouds together with the 6D YSO phase-space distribution could place better constraints on the PTS expansion timescale and dynamical evolution.

\section{Conclusions}
\label{conclusions}

To understand the physical drivers of star formation within the Perseus molecular cloud, we investigated the spatial, kinematic, and temporal imprints left on its YSOs. By analysing 406 Gaia DR3 YSOs, our results suggest that the historical expansion of the PTS likely served as the primary triggering mechanism. Our analysis provides five principal lines of evidence supporting such an expansion-driven framework.

(i) The YSO population exhibits coherent outward motions in 3D space relative to the PTS centre. The full 3D velocity vectors point broadly away from the shell origin. This centre divergent bulk flow demonstrates that the Perseus YSOs share a systematic expansion, consistent with the macroscopic bulk motions and spatial gradients recently reported in \citet{Wang+2022ApJ}.

(ii) Ages decrease with 3D distance to the PTS centre. The age-distance relation shows a significant anti-correlation and a robust negative slope, indicating an outward-younger sequence. This continuous gradient physically links the discrete subgroups previously characterized in \citet{Olivares+2023AandA}.

(iii) Tangential velocity dispersions decrease with 3D distance to the PTS centre. Tangential velocity dispersions and 3D velocity dispersions decline from the inner to the outer region, and robust fits reveal a negative radial trend. The decline of radial dispersion is consistent with the progressive dynamical dispersion growth of older YSO populations under stellar feedback recently demonstrated in \citet{Yang+2025ApJS}, which showed that Class~III YSOs exhibit systematically larger velocity dispersions than Class~II sources in Perseus.

(iv) Directional diagnostics based on infrared evolutionary classifications show that younger YSO classes are more tightly concentrated towards the PTS-centre direction than older populations. This projected spatial segregation provides an independent check consistent with the observed 3D kinematic expansion.

(v) The 3D kinematics of the YSOs and the velocity field of the Perseus molecular cloud exhibit consistent large-scale trends. Rather than a purely radially dispersive pattern, the stellar proper motions show a coherent bulk flow across the sky plane. Concurrently, their line-of-sight velocities display a similar gradient that matches the background $^{12}$CO gas. This shared motion indicates robust kinematic coherence, demonstrating that the distributed stellar population and the natal cloud share a common dynamical flow, quantitatively matching the west-to-east gas velocity gradient from $\sim 4 \text{ km s}^{-1}$ to $\sim 10 \text{ km s}^{-1}$ reported in \citet{Zucker+2018ApJ}.

Quantitatively, for the $406$ YSOs we measured robust negative slopes for both the age-distance relation (\(\frac{\mathrm{d}\,t_{\mathrm{med}}}{\mathrm{d}\,d_{\mathrm{PTS}}}
= -0.05~\mathrm{Myr\,pc^{-1}}\)) and the tangential velocity dispersion–distance trend (\(\frac{d\,\sigma_t}{d\,d_{\rm PTS}} \;=\; -0.016~{\rm km\,s^{-1}\,pc^{-1}}\)). Collectively, the concordant radial trends in age and kinematics, along with the projected directional alignment and the 3D coherence with the molecular gas, favour an expansion-driven interpretation. This scenario posits that the PTS expansion triggered star formation at larger radii, which is both younger and less dynamically dispersed in Perseus, while the inner populations are correspondingly older and more dynamically dispersed. Furthermore, the large-scale morphology of the Perseus molecular cloud itself provides critical structural evidence. The cloud exhibits a pronounced bent or arc-like geometry that opens away from the PTS centre, consistent with being swept up and shaped by an expanding spherical shockwave \citep{Zucker+2021ApJ}. This gas morphology aligns spatially with the kinematic expansion pattern we observe in the stellar population. Although the current data do not uniquely constrain the shell geometry, energy source, or detailed chronology, the observed kinematic and temporal trends provide new evidence for an expansion-driven scenario.

\begin{acknowledgements}

The authors would like to thank the anonymous referee, whose comments have improved the content and presentation of this paper. This research is supported by Guangxi Natural Science Foundation under Grant No.2024GXNSFBA010436 and Guangxi Key Research and Development Program (Guike FN2504240040). ZJL acknowledges support by NSFC grant No.12563005, 12494571 and the National Key R\&D Program of China (Grant Nos. 2024YFA1611704, 2024YFA1611700). This research is also supported by Guangxi Qingmiao Talent Support Program, Bagui Scholars Programme (W.X.-G., GXR-6BG2424001), Guangxi Talent Program (``Highland of Innovation Talents”). This work has made use of data from the European Space Agency (ESA) mission \textit{Gaia} (\url{https://www.cosmos.esa.int/gaia}), processed by the \textit{Gaia} Data Processing and Analysis Consortium (DPAC; \url{https://www.cosmos.esa.int/web/gaia/dpac/consortium}). Funding for the DPAC has been provided by national institutions, in particular the institutions participating in the \textit{Gaia} Multilateral Agreement.

\end{acknowledgements}

%
\bibliographystyle{aa} 
\bibliography{reference} 

\begin{appendix}
\nolinenumbers
\twocolumn

\section{The properties of the 406 Perseus YSOs}
\label{app_406YSOs}

\begin{table*}[!ht]
\renewcommand{\arraystretch}{1.3}
\caption{Properties of the 406 Perseus YSOs.}
\label{tab:mr-perseus-pts}
\centering
\scriptsize 
\setlength{\tabcolsep}{9pt} 
\begin{tabular}{lcccccccccc}
\hline\hline
Group & source ID & RA & Dec & $d_{\mathrm{PTS}}$ & $d$ & $A_{V}$ & $G$ & mean $U$ & mean $V$ & mean $W$ \\
 & & [deg] & [deg] & [pc] & [pc] & [mag] & [mag] & [km s$^{-1}$] & [km s$^{-1}$] & [km s$^{-1}$] \\
\hline
Heleus & 120101686623435648 & 56.050547 & 30.147747 & 182.813 & 399.662 & 0.810 & 16.678 & 8.732 & 21.847 & 2.254 \\
Heleus & 120114846403228928 & 55.953744 & 30.225724 & 180.303 & 397.153 & 1.155 & 14.487 & 8.691 & 21.841 & 2.327 \\
Gorgophone & 120189265300085760 & 55.548166 & 30.609052 & 82.383 & 298.671 & 1.645 & 15.712 & 3.577 & 21.067 & -4.830 \\
Gorgophone & 120329388609121536 & 54.172755 & 30.635572 & 79.862 & 296.561 & 2.110 & 16.321 & 3.786 & 20.999 & -1.329 \\
Gorgophone & 120447379950667008 & 55.085974 & 30.986426 & 83.424 & 299.660 & 0.885 & 15.034 & 3.959 & 21.398 & -1.090 \\
Gorgophone & 120460024334304128 & 54.553059 & 31.074923 & 87.047 & 303.427 & 1.206 & 16.108 & 4.255 & 20.988 & -1.701 \\
Gorgophone & 120463116710737920 & 54.875437 & 31.110552 & 82.388 & 298.593 & 1.576 & 16.399 & 3.687 & 21.325 & -2.368 \\
Gorgophone & 120469404542851712 & 54.837890 & 31.255916 & 71.897 & 287.832 & 0.499 & 17.401 & 3.655 & 21.253 & -1.972 \\
Gorgophone & 120898488955384576 & 52.000456 & 30.146319 & 55.575 & 272.342 & 6.323 & 16.087 & 3.154 & 19.513 & -5.961 \\
Gorgophone & 120914534953167616 & 51.922879 & 30.337933 & 72.900 & 289.708 & 3.836 & 14.285 & 1.458 & 20.213 & -4.431 \\
\hline
\end{tabular}
\tablefoot{Here, we list the main properties of the Perseus YSOs. For each of the 406 Perseus YSOs used in this work, we provide the kinematic and positional group label, Gaia DR3 source identifier, equatorial coordinates at J2016, the 3D distance from the PTS centre ($d_{\mathrm{PTS}}$), the parallax-based distance ($d$), the median line-of-sight visual extinction ($A_V$), the Gaia G-band apparent magnitude, and the ICRS Cartesian velocity components ($U, V, W$). The full 406 Perseus YSO catalogue table is available in electronic form through CDS.}
\end{table*}

Table~\ref{tab:mr-perseus-pts} summarizes the basic properties of the 406 Perseus YSOs used in the age and kinematic analysis in the paper. The complete table is available in machine-readable form at CDS.

\section{Infrared YSO classifications}
\label{app_infrared_YSO}

\begin{table*}
\renewcommand{\arraystretch}{1.3}
\caption{Infrared YSO classifications.}
\label{tab:ir-yso-sample}
\centering
\begin{tabular}{lccc}
\hline\hline
Name & YSO Class & RA & Dec \\
 & & [deg] & [deg] \\
\hline
SSTc2d J034427.9+322718  & II   & 56.116250 & 32.455000 \\
SSTc2d J034439.1+322008  & II   & 56.162917 & 32.335556 \\
SSTc2d J034344.4+314309  & Flat & 55.935000 & 31.719167 \\
SSTc2d J034548.2+322412  & II   & 56.450833 & 32.403333 \\
SSTc2d J034443.0+313733  & II   & 56.179167 & 31.625833 \\
WISE J034707.58+313514.8 & II   & 56.781580 & 31.587437 \\
SSTc2d J034249.1+315011  & II   & 55.704583 & 31.836389 \\
SSTc2d J034412.9+320135  & 0+I  & 56.053750 & 32.026389 \\
SSTc2d J034411.6+320313  & II   & 56.048333 & 32.053611 \\
SSTc2d J034423.6+320152  & II   & 56.098333 & 32.031111 \\
\hline
\end{tabular}
\tablefoot{The table lists the infrared YSO classifications used to define the IR-selected sample. For each member we provide their designation, YSO Class, and sky coordinates. RA and Dec are equatorial coordinates at J2000, in units of degrees. The machine-readable version (MRT) of this table is available at CDS.}
\end{table*}

Table~\ref{tab:ir-yso-sample} lists the infrared-based YSO evolutionary classifications compiled from previous surveys. The complete catalogue is available in machine-readable form at CDS.

\section{Bootstrap robustness of age-distance relation and tangential velocity dispersion-distance relation}
\label{Robustness}

\begin{figure}
\centering
\includegraphics[width=\linewidth]{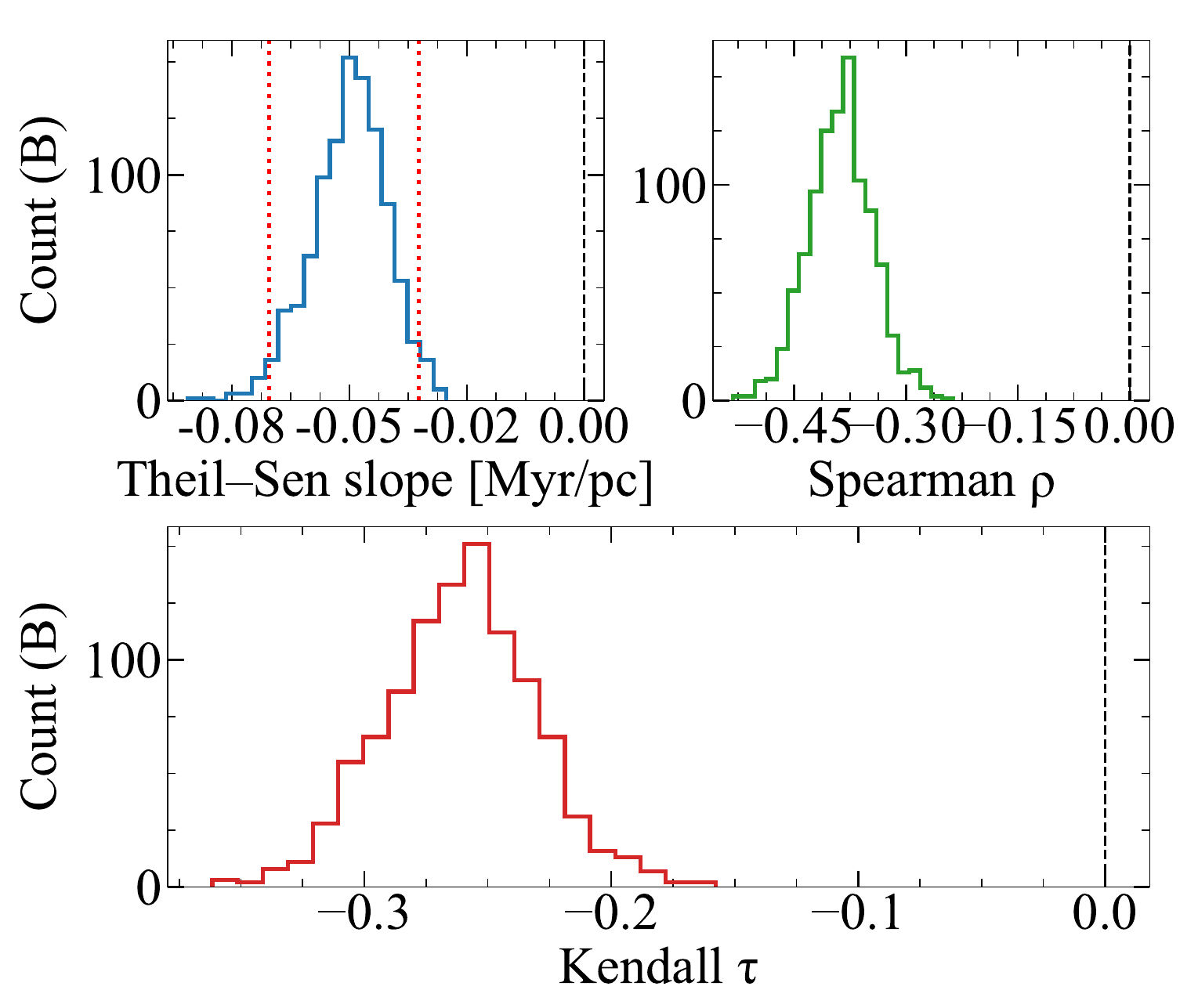}
\caption{Bootstrap robustness check for the individual-star age-distance trend. We resampled the 406 paired measurements $(d_{\mathrm{PTS}}, t_{\mathrm{med}})$ with replacement ($B=1000$) and recomputed the Theil-Sen slope (top left) and rank correlations (top right and bottom). The dotted red lines mark the 95\% percentile interval of the slope distribution.}
\label{fig:robust_slope_bootstrap}
\end{figure}

\begin{figure}
\centering
\includegraphics[width=\linewidth]{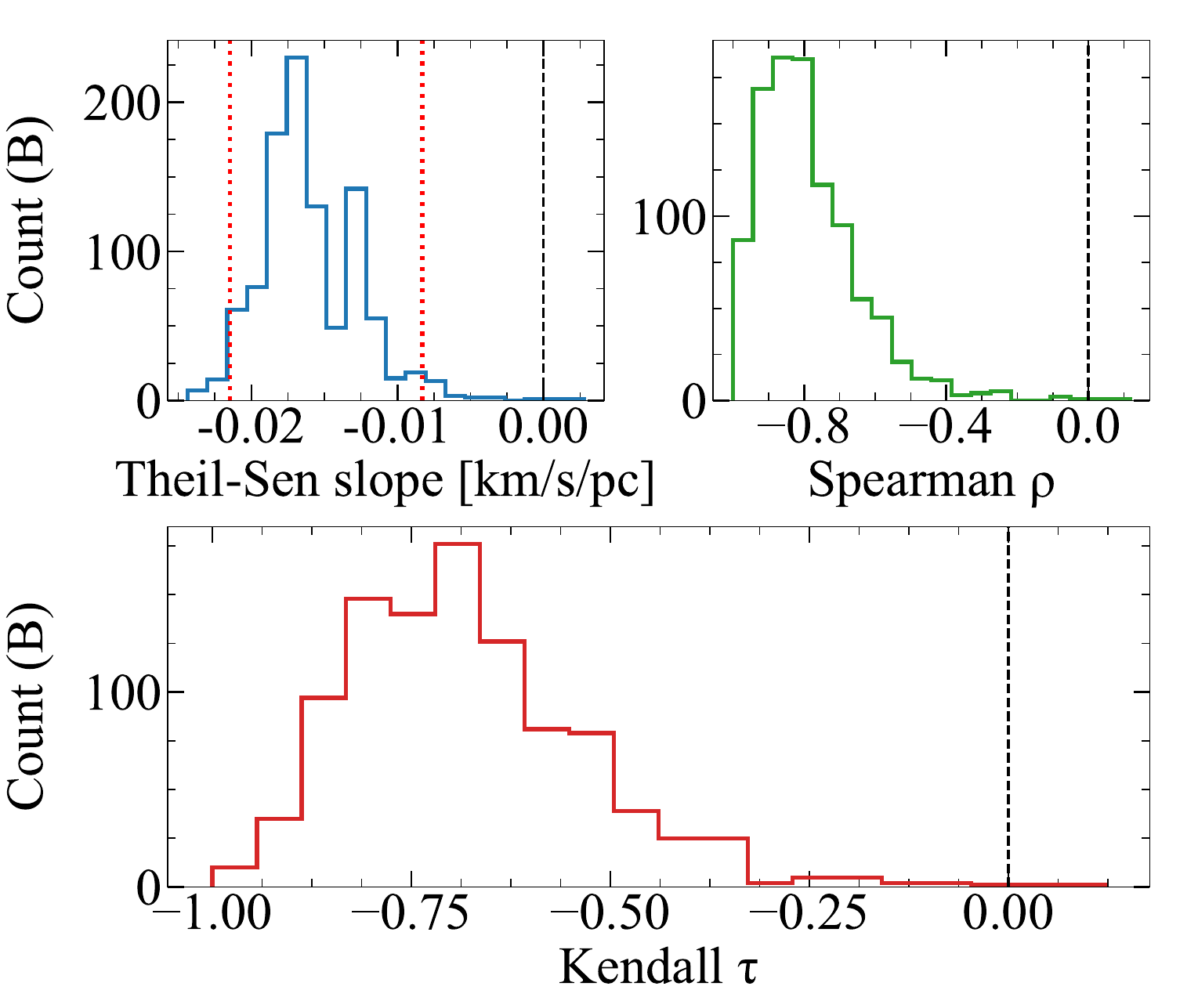}
\caption{Bootstrap robustness check for the $\sigma_t$--distance trend. We bootstrap-resampled the binned points $(d_{\mathrm{PTS}},\sigma_t)$ ($B=1000$) and recomputed the Theil-Sen slope (top left) and rank correlations (top right and bottom). The dotted red lines indicate the 95\% percentile interval of the slope distribution.}
\label{fig:trend_sigma_t_bootstrap}
\end{figure}

As a distribution-free robustness check, we bootstrap-resample the 406 paired measurements $(d_{\mathrm{PTS}}, t_{\mathrm{med}})$ and recompute the Theil-Sen slope (together with rank-based association statistics) for each resample. The resulting sampling distributions are shown in Fig.~\ref{fig:robust_slope_bootstrap}, confirming a consistently negative age--distance slope (95\% interval entirely below zero). 

We further assess the robustness of the $\sigma_t$--$d_{\mathrm{PTS}}$ trend by bootstrap-resampling the binned points $(d_{\mathrm{PTS}},\sigma_t)$ and recomputing the Theil-Sen slope and rank correlations. Figure~\ref{fig:trend_sigma_t_bootstrap} shows the corresponding bootstrap distributions, supporting a stable monotonic trend.

\section{Robustness against PTS centre uncertainties}
\label{app_robustness_joint}

\begin{figure}
\centering
\includegraphics[width=\linewidth]{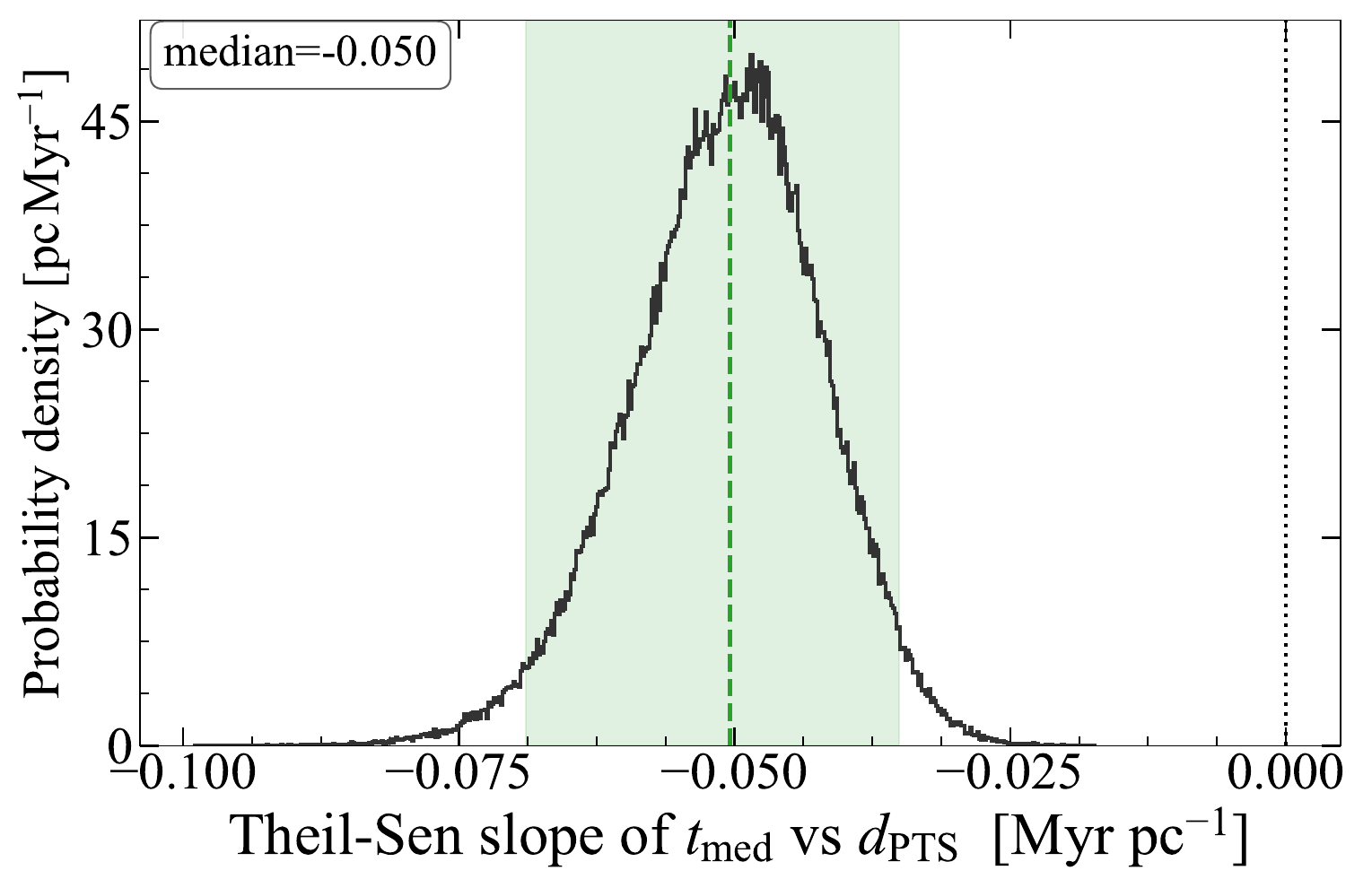}
\caption{Distribution of the Theil-Sen slopes of the $t_{\rm med}$-$d_{\rm PTS}$ relation under the joint uncertainty of the adopted PTS centre. The x-axis shows the Theil-Sen slope of $t_{\rm med}$-$d_{\rm PTS}$ relation obtained from each perturbed centre realization and bootstrap resampling, while the y-axis shows the corresponding probability density. The dotted black line marks the null value of zero slope, corresponding to no radial gradient in median age. The dashed green line indicates the median slope, and the green shaded area marks the 95\% percentile interval of the slope distribution.}
\label{fig_jointage}
\end{figure}

\begin{figure}
\centering
\includegraphics[width=\linewidth]{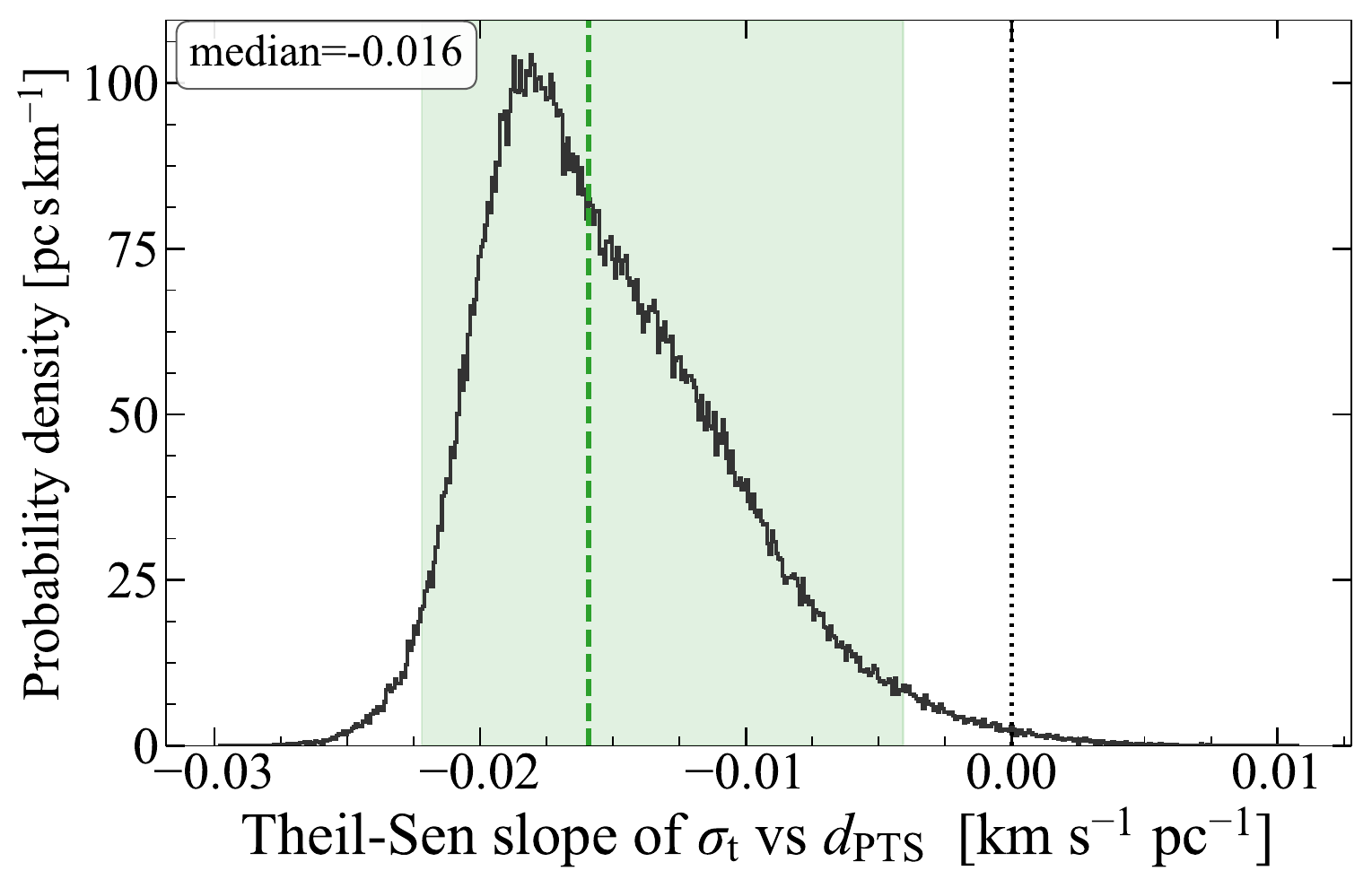}
\caption{Same as Fig.~\ref{fig_jointage} but for the distribution of the Theil-Sen slopes of the $\sigma_t$-$d_{\rm PTS}$ relation under the joint uncertainty of the adopted PTS centre.}
\label{jointvdisp}
\end{figure}

In our primary analysis, we adopt the geometric centre of the PTS from 3D dust mapping, $(X,Y,Z)=(-190,\,65,\,-84)\,\mathrm{pc}$. To quantitatively test the sensitivity of our age-distance gradient to the uncertainty in this reference point, we perform a nested Monte Carlo perturbation test.

We simulate $N_{\mathrm{center}}=500$ independent centre realizations. In each realization, we perturb the PTS centre in heliocentric Galactic Cartesian coordinates by drawing $\Delta X$, $\Delta Y$, and $\Delta Z$ from $\mathcal{N}(0,\sigma_{\rm c}^2)$ with $\sigma_{\rm c}=15\,\mathrm{pc}$. We adopt this value as a conservative scale for the centre-position uncertainty of the 3D dust reconstruction at the distance of Perseus. For each perturbed centre, we recompute $d_{\mathrm{PTS}}$ for the YSO sample and perform $N_{\mathrm{boot}}=500$ bootstrap resamples with replacement. For each resample, we estimate the Theil--Sen slope of $t_{\mathrm{med}}$ versus $d_{\mathrm{PTS}}$.

Figure~\ref{fig_jointage} shows that the inferred negative age-distance relation is robust within the plausible geometric parameter space of the PTS centre. The slope distribution remains entirely negative, with a median slope of $-0.050~\mathrm{Myr~pc^{-1}}$.

We apply the same joint perturbation test to the kinematic trend between the tangential velocity dispersion and the PTS-centred distance. For each perturbed centre and each bootstrap realization, we recompute $d_{\mathrm{PTS}}$ for the full YSO sample and reassign stars to the same set of distance bins. We then estimate the intrinsic tangential velocity dispersion $\sigma_t$ in each bin using the heteroscedastic MLE and measure the Theil--Sen slope of $\sigma_t$ versus $d_{\mathrm{PTS}}$. The resulting slope distribution as shown in Fig.~\ref{jointvdisp} remains negative, with a median slope of $-0.016~\mathrm{km\,s^{-1}\,pc^{-1}}$. This indicates that the decrease of $\sigma_t$ with increasing $d_{\mathrm{PTS}}$ is robust against plausible 3D uncertainties in the adopted PTS centre and against finite-sample variability.


\end{appendix}

\end{document}